%% file: ilc-tdr-barish_brau.tex
\documentclass{ws-ijmpa}
\usepackage[super,compress]{cite}
\usepackage{graphicx}
\usepackage{pdfpages}
\usepackage[maybess]{heppennames2} %symbols for physics processes (from CERN)
\newcommand{\gghadrons}{\ensuremath{\upgamma\upgamma \rightarrow \mathrm{hadrons}}\xspace}
\newcommand{\roots}{\ensuremath{\sqrt{s}}\xspace}
\begin{document}
\markboth{Barish and Brau}
{The International Linear Collider}

\catchline{}{}{}{}{}
%
%%%%%%%%%%%%%%%%%%%%%%%%%%%%%%%%%%%%%%%%%%%%%%%%%%%%%%%%%%%%%%%%%%%%

\title{The International Linear Collider}

\author{Barry Barish}

\address{California Institute of Technology\\
Pasadena, California 91125, USA\\
barish@ligo.caltech.edu
}

\author{James E. Brau}

\address{Center for High Energy Physics, University of Oregon\\
Eugene, Oregon 97403-1274, USA\\
jimbrau@uoregon.edu
}

\maketitle

\begin{history}
\received{Day Month Year}
\revised{Day Month Year}
\end{history}

\begin{abstract}
In this article, we describe the key features of the recently completed technical design for the International Linear Collider (ILC), a 200-500 GeV linear electron-positron collider (expandable to 1 TeV) that is based on 1.3 GHz superconducting radio-frequency (SCRF) technology. The machine parameters and detector characteristics have been chosen to complement the Large Hadron Collider physics, including the discovery of the Higgs boson, and to further exploit this new particle physics energy frontier with a precision instrument.   The linear collider design is the result of nearly twenty years of R\&D, resulting in a mature conceptual design for the ILC project that reflects an international consensus.  We summarize the physics goals and capability of the ILC, the enabling R\&D and resulting accelerator design, as well as the concepts for two complementary detectors.   The ILC is technically ready to be proposed and built as a next generation lepton collider, perhaps to be built in stages beginning as a Higgs factory.

\end{abstract}

\ccode{PACS numbers:13.66-a}

\tableofcontents

\section{Introduction}

\input intro.tex

\section{Physics Purpose and Goals}

\input physics.tex

\section{The Collider}

\input collider.tex

%\includepdf[pages={1-19}] {accelerator/WS_Accelerator_ver2.pdf}

\section{The ILC Detectors}

\input detectors.tex

\section{Conclusion}

\input conclusion.tex

\section*{Acknowledgments}

\input acknow.tex

%\section*{References}

%\begin{thebibliography}{000} %for 3 digits
%\begin{thebibliography}{00}  %for 2 digits

\end{document}

%% file: intro.tex
The International Linear Collider (ILC) should play a critical role in the future
development of elementary particle physics.  
This 200-500 GeV (extendable to 1 TeV) centre-of-mass energy, high luminosity,
linear electron-positron collider, based on 1.3 GHz superconducting radio-frequency (SCRF)
accelerating technology, offers new powerful physics capabilities
including precision studies of 
the Higgs boson discovered in 2012 at the Large Hadron Collider at CERN.\cite{atlas-higgs,cms-higgs}

With the development of particle colliders fifty years ago, particle physics
entered a new era.
The huge increase in center-of-mass
collision energy made possible by colliders,
compared to the classical particle accelerators where a beam of particles struck a fixed target, proved revolutionary for the science.   Since that time, three generations of colliding beam machines have been built and exploited,
going to higher and higher energies, and in each case, benefiting from the complementarity of hadron and electron-positron colliders.

The first generation of electron-positron colliders included SPEAR at SLAC which, along with the Brookhaven Alternating Gradient Synchrotron hadron accelerator, was responsible for discovery 
of the J/psi (J/$\psi$) particle. This initiated a fantastic period of discovery in particle physics, where most importantly charmed particles were discovered at SPEAR, as well as the heaviest lepton, the tau.   The next generation electron-positron colliders, especially PETRA at DESY, were responsible for the discovery of the gluon, a key for quantum chromodynamics, the basis for understanding the strong interaction.  Finally, the last generation electron-positron collider was LEP at CERN that gave the brilliant confirmations of the Standard Model of particle physics through a long series of precision measurements and tests.  The combination of lepton and hadron colliders has proven to be extraordinarily powerful in probing each high energy frontier of particle physics. 

 Over the past decade, studies conducted worldwide developing future priorities for particle physics have established a consensus that a high energy lepton collider 
is the top priority in the field.  This is the role of the linear collider.
There is every reason to expect that much like the LHC, the future linear collider when built will be one of the world's premier science facilities.

The ILC design is the result of nearly twenty years of R\&D. A 
worldwide international collaboration coordinated by the Global Design
Effort (GDE) under a mandate from the International Committee for Future Accelerators (ICFA)
has developed the design since 2005.
This effort has successfully culminated in the publication of a Technical Design Report (TDR)\cite{tdr},
describing the R\&D which has overcome several high-risk challenges in the collider
technology.
The TDR also presents detailed designs of two detectors developed by large international teams
as a result of intense detector R\&D. This paper highlights the key features of the ILC TDR.

%% file: physics.tex
Particle physicists have successfully established a Standard Model of the fundamental particles and their interactions;
it has passed numerous tests by many experiments.\cite{StandardModel}
The Standard Model
includes an explanation for the origin of electroweak symmetry breaking 
and the generation of the mass of the fundamental particles
through the Higgs mechanism involving a Higgs boson. A candidate for the Standard Model Higgs boson has been discovered at the LHC with properties, 
at the current level of precision,
expected
by the theory.  But many details of this particle's properties, and the presence or absence 
of other anticipated new particles, remain to be discovered to complete
the theoretical picture.

While the Standard Model has been very successful, it does not explain a number of fundamental features of nature,
such as: 
\begin{romanlist}
\item does the particle found at the LHC fully explain the origin of electroweak symmetry breaking as well as the origin of mass of all fundamental particles?
\item if the particle found at the LHC is the Standard Model Higgs boson, why is its mass what it is
and not much larger?  This is known as the hierarchy problem. 
\item what is the source of the neutral, weakly interacting matter (dark matter) that appears to dominate the matter of the universe?
\item why is the atomic matter of the universe today composed of matter but little anti-matter?
\end{romanlist}

The capabilities of the ILC make it an ideal tool to explore and find answers to these questions. It  will begin by
making very precise measurements of the properties of the LHC-discovered Higgs boson. Are these properties
exactly consistent with the expectations of the Standard Model?  The physics program will continue with
precise measurements of other Standard Model particles only possible in the experimental environment 
provided by the electron-positron collider. Along the way, searches for new particles will be pursued.
Table \ref{tab:ILCprogramX} presents possible center-of-mass energies for ILC operation along with the
physics goals at each operating point.\cite{tdr}

   The Higgs potential provides an important test of the Standard Model interpretation of the 
   candidate Standard Model Higgs particle.  The Standard Model potential gives rise to triple
   and quartic self-couplings of the Higgs, well defined by the mass of the Higgs itself.
   The ILC will measure the triple self-coupling term.\cite{tdr-2}

The collider evolution is being planned in a `staged' scenario to realize this
physics program.
Under this `staged' plan, the first phase of the ILC would be a Higgs factory with a center-of-mass energy of approximately 250 GeV.  Following a period of
operation of the 250 GeV machine, it would be upgraded in stages, to
a center-of-mass energy of $\sim$500 GeV, which is the baseline energy of the overall project.  The project would ensure the eventual extendibility to 1 TeV,
which could be reached in the latter stages of the program.

\begin{table}[t] 
%\begin{center}
\tbl{Major physics processes to be studied by the ILC at various 
energies.   The  table indicates the 
various Standard Model reactions that will be accessed at increasing 
collider energies, 
and the major physics goals of the study of these reactions. A reaction 
listed at a given energy will  
be studied at all higher energies.}
{\begin{tabular}{rccc}
Energy             &   Reaction  &  Physics Goal     
 \\  \hline \hline
91~GeV            &       $e^+e^-  \to Z$     &      
  ultra-precision electroweak        \\   \hline
160~GeV            &       $e^+e^-  \to WW$     &
ultra-precision $W$ mass         \\   \hline
250~GeV         &    $e^+e^-  \to Z h$     &      precision Higgs
couplings           \\ 
                          \hline
350--400~GeV         &      $e^+e^-  \to t\bar t$        &  
  top quark mass and  couplings                \\
  &    $ e^+e^-  \to WW$     &     precision $W$ couplings            \\ 
                       &     $ e^+e^-  \to \nu\bar\nu h $   & 
 precision Higgs couplings     \\   \hline
500~GeV         &      $e^+e^-  \to f\bar f   $       &   
 precision search for $Z'$      \\ 
    &  $ e^+e^-  \to t\bar t h$    &      Higgs coupling to top      \\
                        &      $e^+e^-  \to Z h h $       &     Higgs
                        self-coupling            \\
                       &   $   e^+e^- \to  \tilde \chi \tilde \chi$  &
                       search for supersymmetry   \\  
                       &    $  e^+e^- \to AH, H^+H^-  $  &
                       search for extended Higgs states   \\  \hline
700--1000~GeV &        $ e^+e^-  \to \nu\bar \nu hh $ & 
        Higgs self-coupling     \\ 
                          &    $e^+e^-  \to \nu\bar \nu VV$ &  composite
                          Higgs sector  \\ 
                          &    $e^+e^-  \to \nu\bar \nu t\bar t$ &
                          composite Higgs and top  \\
                          &    $e^+e^-  \to \tilde t \tilde{ t}^* $ &
                        search for supersymmetry  \\
                           \hline
\end{tabular}
\label{tab:ILCprogramX}}
%\end{center}
\end{table}

The announcement by the ATLAS and CMS Collaborations of the discovery of a particle with mass of about 125 GeV in July, 2012,\cite{atlas-higgs,cms-higgs} followed a long period of speculation on the indications that such a particle was needed
to complete the Standard Model of particle physics.\cite{higgs}  This newly discovered particle does, indeed, carry many of
the properties of the postulated Higgs boson of the Standard Model.  
The LHC experiments are continuing to refine measurements of its properties, but the LHC precision will ultimately 
be limited.
The questions listed above must be addressed and the ILC is the ideal complement to the LHC for this.

\begin{figure}[h]
%\centerline{\includegraphics[width=12.8cm]{images/physics/FIG-2_6.png}}
\centerline{\includegraphics[width=12.8cm]{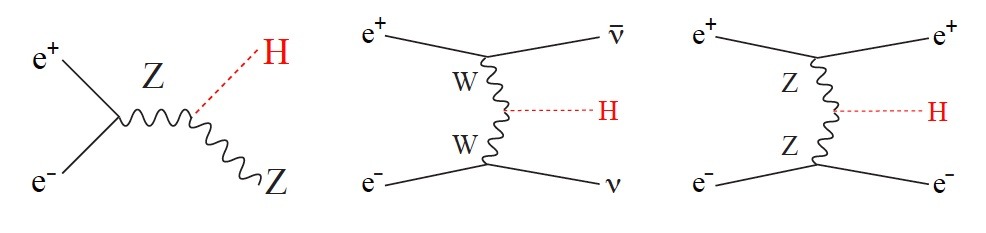}}
\caption{Feynman diagrams for the three major Higgs production processes at
the ILC: $e^+e^- \rightarrow Zh$ (left), $e^+e^- \rightarrow \nu \overline{\nu} h$ (center),
and $e^+e^- \rightarrow e^+e^- h$ (right). \label{higgs-feynman}}
%\end{figure}
\parbox{.3in}{}
%\begin{figure}[h]
\centerline{\includegraphics[width=8.8cm]{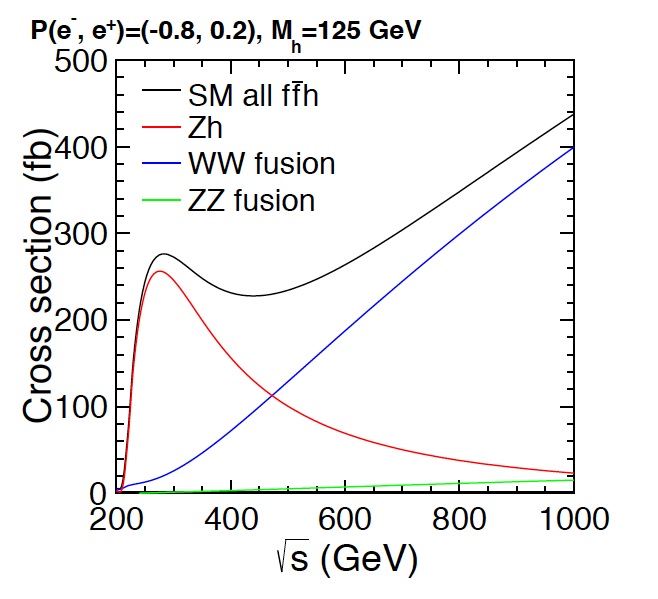}}
\caption{Production cross sections for Higgstahlung ($e^+e^- \rightarrow Zh$), WW fusion
($e^+e^- \rightarrow \nu \overline{\nu} h$), and ZZ fusion ($e^+e^- \rightarrow e^+e^- h$)
as a function of the center-of-mass energy for $m_h = 125$ GeV and beam polarization
(P$_{e^-}$, P$_{e^+}$) = (-0.8, +0.2). \label{higgs-cross-sec}}
\end{figure}

The three major Higgs production processes at the ILC are illustrated in Fig. 
\ref{higgs-feynman}. Fig.~\ref{higgs-cross-sec} presents the energy dependence of the cross
sections for these processes.  At lower energy, the Higgs-strahlung ($e^+e^- \rightarrow Zh$)
process is dominant, and as the energy increases
Higgs-strahlung weakens and the
 WW fusion process increases and dominates Higgs production.
The ILC's ability to make precision measurements from these processes results
from its
providing well defined initial states, a clean environment, and good signal-to-noise levels
prior to any event selection cuts.

\begin{figure}[htbp]
\parbox{2.4in}{
\begin{center}
\includegraphics[width=2.4in]{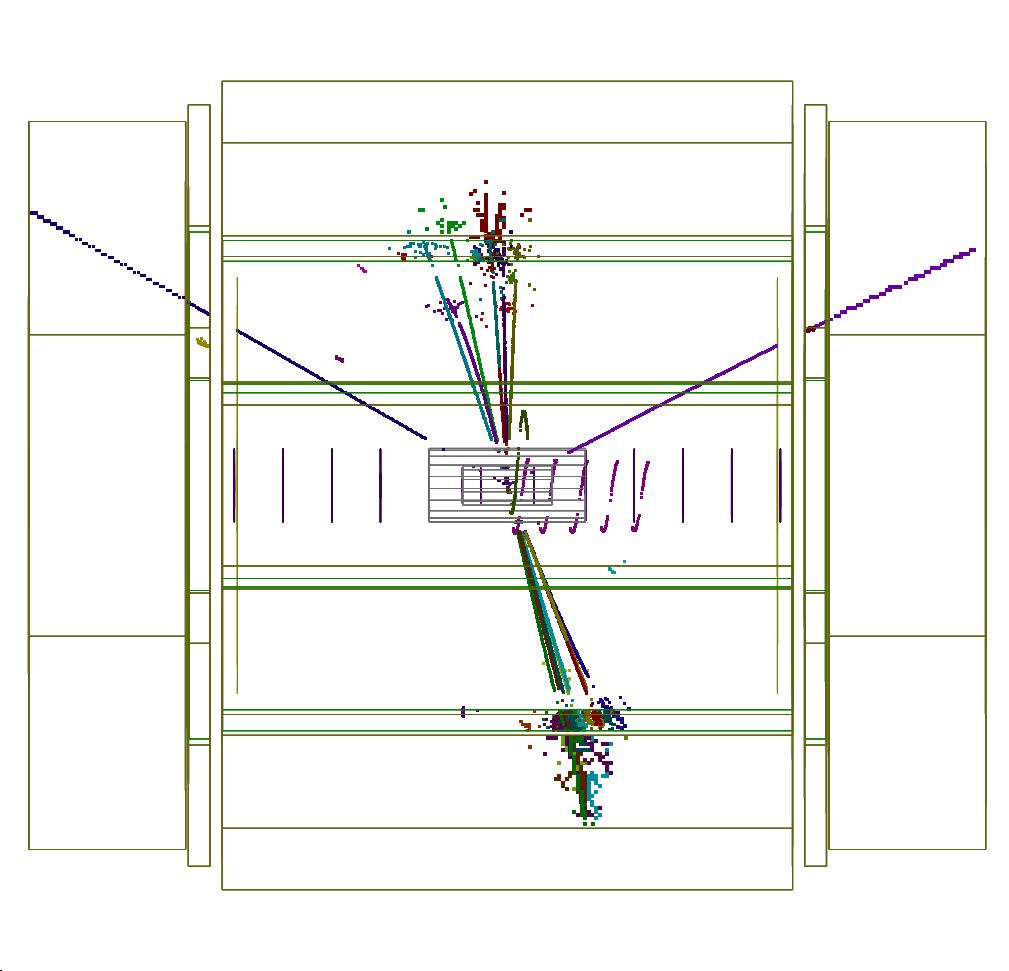}
\end{center}
%\caption{ 
%}
}~~~~~\parbox{2.8in}{
\begin{center}
\includegraphics[width=2.4in]{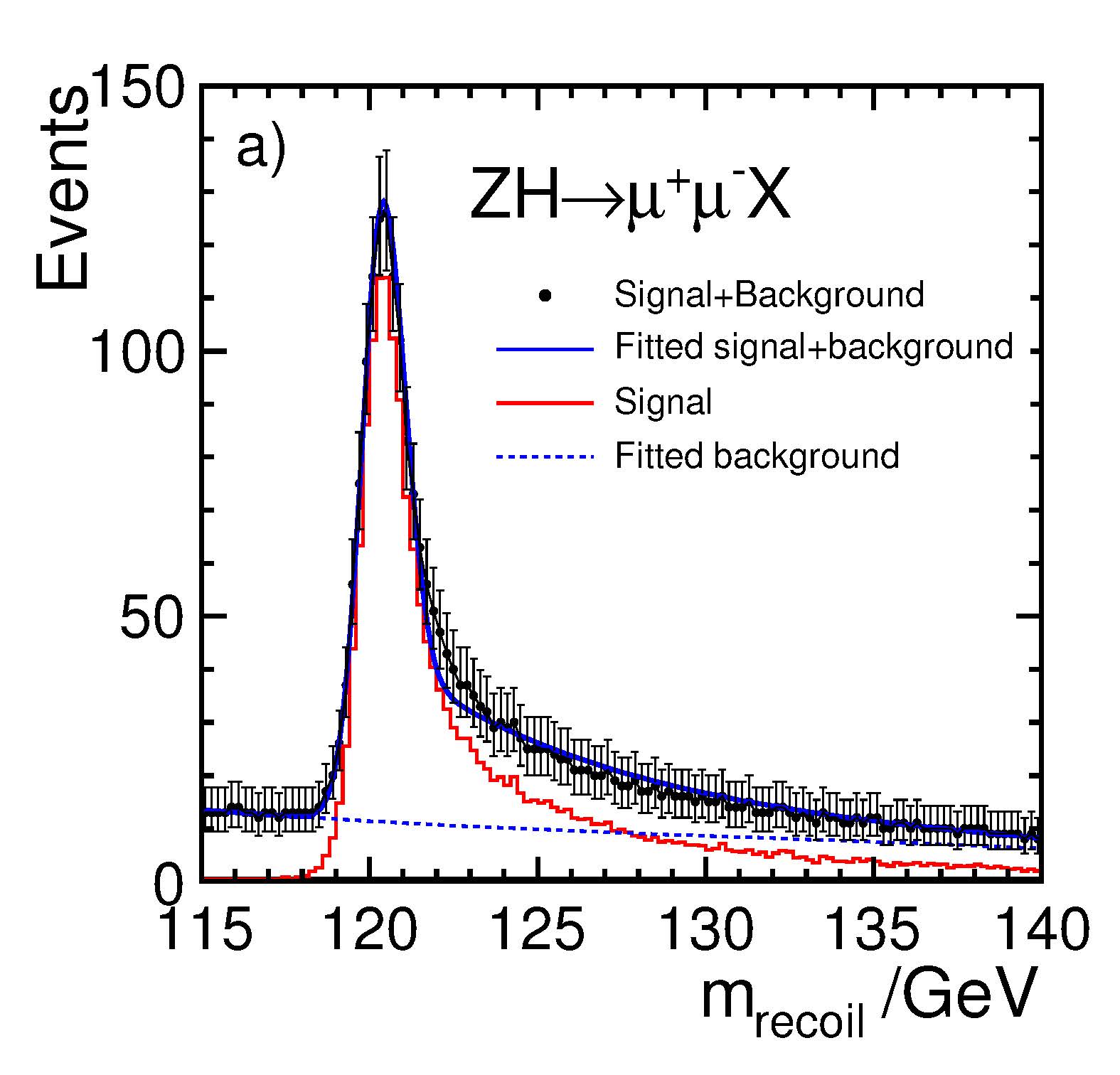}
\end{center}
%\caption{ }
%}
}
\caption{(a) An event of reaction $e^+e^- \rightarrow Zh$, with
$Z \rightarrow \mu^+\mu^-$, $h \rightarrow b\overline{b}$, as it would be
observed in the ILD detector at the ILC. (b) Results of a model independent analysis
of the Higgs-strahlung process $e^+e^- \rightarrow Zh$, in which
$Z \rightarrow \mu^+\mu^-$, as it would be
observed in the ILD detector at the ILC, based on 250 fb$^{-1}$. }
\label{ild-higgs}
\end{figure}

The initial ILC physics program will make use of the peak in the Higgs-strahlung cross section
near center-of-mass energy
250 GeV.  Here, the presence of a clear Z boson makes it possible to tag events
having a recoil mass against the observed Z boson appropriate for the Higgs boson.  
Fig.~\ref{ild-higgs}(a) shows a simulation of
one Higgs-strahlung event where the Higgs
 boson decays to quarks as it would be seen in an ILC detector.
Such events
are detected through the Z decay products without investigating the specifics of the Higgs decay itself.  
Fig.~\ref{ild-higgs}(b) presents a 250 fb$^{-1}$ simulation of the measured recoil mass for the case in which
Z decays to $\mu^+\mu^-$.
In this 
way, rates for all of the decays of the Higgs boson - including decays to invisible
or unusual final states - can be measured with high precision.
At the LHC, such events are very difficult to separate from an overwhelming Standard Model background.
The precision of the ILC {\it model independent} measurement of rates of decay of the Higgs boson to the various 
channels (quarks, leptons, and bosons) will inform the study of whether the Higgs field
alone gives mass to the fundamental particles, or if additional new particles share the 
mass generating role of the Higgs field. There are many highly motivated new physics scenarios that
generate deviations from the Standard Model in the sub-ten percent range.\cite{grw}
The ILC precision will address these scenarios.

%\begin{figure}[h]
%\centerline{\includegraphics[width=8.8cm]{images/detectors/ES/Higgsevent.pdf}}
%\caption{An event of reaction $e^+e^- \rightarrow Zh$, with
%$Z \rightarrow \mu^+\mu^-$, $h \rightarrow b\overline{b}$, as it would be
%observed in the ILD detector at the ILC. \label{ild-higgs}}
%\end{figure}

%\begin{figure}[h]
%\centerline{\includegraphics[width=8.8cm]{images/detectors/FIG_III-6-6-left-new_mh_mmX.pdf}}
%\caption{Results of a model independent analysis
%of the Higgs-strahlung process $e^+e^- \rightarrow Zh$, in which
%$Z \rightarrow \mu^+\mu^-$, as it would be
%observed in the ILD detector at the ILC. \label{recoil}}
%\end{figure}

The program of measurements at 250 GeV will include the measurement of mass and spin
of the Higgs boson.  While the LHC already provides excellent access to these properties,
the ILC brings new, experimentally attractive probes. For example, the recoil mass
measurement illustrated in Fig. \ref{ild-higgs}(b) allows the mass of the Higgs boson
to be determined 
from 250 fb$^{-1}$ at 250 GeV
with a statistical precision of 40 MeV in the $Z \rightarrow \mu^+\mu^-$
channel
and independently to 80 MeV precision in the $Z \rightarrow e^+e^-$ channel.
Combining the two results in an uncertainty of 35 MeV.  The ILC also offers
a complementary test for the spin of the Higgs boson to that of the LHC.
The threshold behavior of the $e^+e^- \rightarrow Zh$ cross section varies
significantly between possible spin and CP values for the Higgs boson.   
The spin 1 possibility is ruled out by observation of $h \rightarrow \gamma\gamma$
at the LHC.  This decay also restricts the charge conjugation value to be positive.
For spin 0, the threshold behavior of the cross section
at the ILC
 is $\beta$ for CP even
and $\beta^3$ for CP odd, while for spin 2 the threshold rise is also 
expected to be $\beta^3$. For larger spins the rise will follow higher powers
of $\beta$. 
Measuring the cross section at three points just above threshold
with 20 fb$^{-1}$ at each point clearly discriminates between these
possibilities.
At energies well above threshold (eg. 350 GeV) the $Zh$ process is dominated by
longitudinal Z production.  This will be used to verify the scalar nature of the
Higgs boson.  The ILC also has the ability to address more complicated possibilities,
such as the case in which the Higgs boson is not a CP eigenstate, but a mixture of
CP even and CP odd.  The unique experimental environment and beam parameters
makes ILC a powerful tool to test these possibilities.\cite{tdr-2}

The inclusive cross section of the $e^+e^- \rightarrow Zh$ process is a direct measure
of the $h$ to $ZZ$ coupling, $g_{hZZ}$.  The partial width of the $h \rightarrow ZZ$ decay
($\Gamma(h \rightarrow ZZ)$) is itself directly proportional to the inclusive
$e^+e^- \rightarrow Zh$ cross section.  Given also the branching ratio of
$h \rightarrow ZZ$ obtained using the recoil method, the total width of the
Higgs boson (expected to be 4 MeV in the Standard Model) can be obtained

$$\Gamma_{tot} = {\Gamma(h \rightarrow ZZ) \over BR (h \rightarrow ZZ)}$$

\noindent with high precision
of about 2\% with 250 GeV (250 fb$^{-1}$) and 500 GeV (500 fb$^{-1}$) data combined.

Following the initial measurements of the Higgs boson at 250 GeV, the program 
will continue at higher energies.  
As the energy reaches 350 GeV, the ILC will cross
a prominent rise in the cross section due to the top quark pair
threshold.  The shape of this rise is precisely predicted by 
perturbative QCD and measurement of its shape will yield a top
quark mass determination with an accuracy of 100 MeV,
a precision that is not reachable at the LHC.
This determination is a crucial input to fundamental physics predictions,
such as those of grand unification.
In addition, measurements of the details of the top pair events near
threshold and above will lead to precision measurements critical
in the theory of electroweak symmetry breaking.
With the unique beam polarization available to the ILC,
sensitivity to new physics associated with composite Higgs bosons,
for example, will be achieved. 

In the center-of-mass energy region of 350-400 GeV,
the reaction $e^+e^- \rightarrow W^+W^-$ provides
a very sensitive probe to modifications of the Standard Model 
at high energy.
As the energy is increased further, the WW fusion process
($e^+e^- \rightarrow \nu \overline{\nu} h$)  grows, yielding
a measurement of the $hWW$ coupling, another crucial
addition to the precision Higgs boson studies.

At the full design energy of the ILC, 500 GeV, the WW fusion process (see Fig. \ref{higgs-cross-sec})
grows ever stronger,
providing an absolute normalization of the Higgs coupling strengths.
The ILC is also sensitive to the Higgs self-coupling at 500 GeV.
At this energy, precision studies of the two-fermion reactions
$e^+e^- \rightarrow f \overline{f}$ probe with great sensitivity
high mass vector resonances, new fermion interactions, and quark
and lepton compositeness.  Searches for new particles will also be possible,
such as color-singlet supersymmetric particles and states associated with an
extended Higgs sector.  These are searches that are very difficult at the LHC.

The study of pair production of quarks, leptons, and W and Z bosons,
at the ILC will provide critical new knowledge on possible
new interactions at higher mass scales.  The detailed study of the
properties of the W boson and the top quark will add to our existing
precision measurements of the Z boson from $e^+e^-$ colliders, which
provides a significant level of current Standard Model knowledge.
For example, the top quark mass measurement of 100 MeV precision from a 
threshold scan, as described above, is an important parameter
for accurate theoretical calculations.
%The top quark mass will be measured directly through a threshold scan
%that is not 
%possible at the LHC.  The precise value of the top quark mass is an
%important parameter for accurate theoretical calculations.  
%The top quark
%also provides a possible indicator of the presence of composite structure
%in the Higgs particle which could show up in the measurement of the
%strong coupling of the top to the Higgs field.  
The precision of top and W
measurements are summarized in Table \ref{tab:TopSummary}

\begin{table}[h]    %[p] 
%\begin{center}
\tbl{Key precision Standard Model measurements of the top quark and the W boson that will be achieved with
the ILC}
{\begin{tabular}{lccl}
Topic          &   Parameter & Accuracy  $\Delta X/X$   
 \\  \hline \hline
Top           &    $  m_t   $  &      0.02\%   &   $\Delta m_t =
30$~MeV,   threshold scan \\ 
                     &     $\Gamma_t  $&   2.\%     &    \\ 
                      &    $\tilde F^\gamma_{1V} $      &    0.2\%     &      500
                      GeV\\ 
                      &    $\tilde F^Z_{1V} $      &    0.3\%     &     \\
                      &    $\tilde F^Z_{1A} $      &    0.5\%     &     \\
                      &    $\tilde F^\gamma_{2V} $      &    0.3\%     &     \\
                      &    $\tilde F^Z_{2V} $     &    0.6\%     &
                      \\ \hline
$W$          &    $  m_W   $  &      0.004\%   &   $\Delta m_W =
3$~MeV,   threshold scan \\ 
                     &    $ g_1  $&   0.16\%     &   500~GeV \\ 
                      &    $\kappa_\gamma $      &    0.03\%     &     \\ 
                      &    $\kappa_Z  $      &    0.03\%     &     \\  
                      &    $\lambda_\gamma $      &    0.06\%     &     \\ 
                      &    $\lambda_Z  $      &    0.07\%     &     \\
\end{tabular}
\label{tab:TopSummary}}
%\end{center}
\end{table}

As the energy is increased further, 
to the upgraded 1 TeV level,
the precise  Yukawa couplings of the Higgs boson to top quarks
is measured, and the precision of the Higgs boson self-coupling 
measurement reaches 20\%.  
The program of searches for new exotic particles continues,
and probes of strongly-interacting or composite models of the Higgs boson
are available.
The top quark
provides a possible indicator of the presence of composite structure
in the Higgs particle which could show up in the measurement of the
strong coupling of the top to the Higgs field.  

Table \ref{tab:HiggsSummary} summarizes the Higgs precision
that the ILC experiments expect to achieve 
based on the TDR collider parameters. \cite{tdr}
These are model independent, unlike the model dependent
analysis that is frequently employed to test standard model couplings
such as the Higgs coupling parametrization proposed by the LHC Higgs Cross Section Working Group (HXSWG) 
has proposed a series of benchmark Higgs coupling parameterizations
.\cite{hxswg}
% A. David et al. LHC HXSWG interim recommendations to explore the coupling structure of a
%Higgs-like particle. 2012, 1209.0040.
%S. Dittmaier, C. Mariotti, G. Passarino, R. Tanaka, et al. Handbook of LHC Higgs Cross Sections:
%1. Inclusive Observables. 2011, 1101.0593.
When such assumptions are made for the ILC precisions significantly better accuracy can be obtained.

\begin{table}[b]    %[p] 
%\begin{center}
\tbl{Estimates for ILC
measurements of Standard Model Higgs boson ($m_h = 125$ GeV) properties. 
These analyses require no significant model-dependent
assumptions. Measurement accuracies are quoted
for ILC event samples of  250 fb$^{-1}$ at 250~GeV, 500 fb$^{-1}$ at
500~GeV, amd  1000 fb$^{-1}$ at 1000~GeV, with electron/positron 
polarization of 80\%/30\% at the first two energies and 80\%/20\% at the
third
energy.}
{\begin{tabular}{lccl}
Topic          &   Parameter & Accuracy  $\Delta X/X$   
 \\  \hline \hline
Higgs            &   $  m_h $   &      0.03\%   &   $\Delta m_h =
35$~MeV,   250 GeV \\ 
                &    $  \Gamma_h  $ &   2.\%     &     250 GeV and 500
                     GeV \\                     
                    &   $  g(hWW)  $  &  0.3\%    &        \\ 
                     &    $ g(hZZ)  $    &  0.35\%    &        \\ 
                     &    $ g(hb\bar b)  $ &  1.1\%    &        \\ 
                     &    $ g(hc \bar c)  $ &  2.1\%    &        \\ 
                     &    $ g(hgg)  $ &  2.3\%    &        \\ 
                     &    $ g(h\tau^+\tau^-)  $ &  2.0\%    &        \\ 
                     &   $  BR(h\to \mbox{ invis.})  $ &   0.05\%  &
                     \\ 
                      &    $ g(ht\bar t)  $  &  4.5\%    &    1000 GeV    \\ 
                      &    $ g(hhh)  $  &  20.\%    &     \\ 
                     &   $  g(h\mu^+\mu^-)  $ &  16.\%    &        \\  

\end{tabular}
\label{tab:HiggsSummary}}
%\end{center}
\end{table}

New particles associated with the Higgs field, dark matter, or other open issues
in particle physics are each important possible future discoveries.
The low mass of the Higgs boson is mysterious, since radiative corrections
to its mass will drive it to a much larger mass, up to the Planck scale,
unless there is new physics
within the TeV regime.  This is known as the `hierarchy problem.'
The ILC will search for particles associated with this new physics.
Many of the hypothesized new particles are only weakly interacting, 
making their discovery at the LHC very difficult because of
their low production rates relative to the rate for strongly interacting particles, and
by the large backgrounds at the hadron collider.  These rate and background issues
are mostly eliminated at the ILC, allowing experiments to discover and
identify, or excluded unambiguously, particles with mass at least
up to the ILC beam energy.

Among the models for new physics\cite{tdr-2} which introduce particles 
designed to solve the `hierarchy problem' is the popular supersymmetry
%There are many models for new physics, largely motivated to solve the
%`hierarchy problem,'
% in which additional particles
%are required. Among such models is the popular supersymmetry
theory based on a symmetry of space and time, and
 between matter (quarks and fermions)
and force-carrying particles.  
This symmetry includes superpartners for all known fundamental particles,
which introduce additional radiative corrections to the Higgs boson
mass which can cancel the divergences that drive the mass to large scales,
solving the `hierarchy problem'.
Such new particles are a prominent focus of the LHC physics programs.
No evidence has yet appeared in the 8 TeV LHC data, but the 14 TeV data
will significantly extend the superpartner reach. The LHC is particularly
sensitive to colored, strongly interacting superpartners.

Supersymmetry introduces matter-like
Higgs particles, which may be difficult to detect at the LHC.  
The existence of such particles is also motivated by their potential
role in explaining the dark matter content of the universe.
They are
produced rarely and release very little energy.  In contrast to the LHC, they are
easily detectable at the ILC, and, in addition, the ILC can measure their quantum
numbers unambiguously and determine their couplings to the percent 
level.

The ILC has the ability to respond to LHC discoveries of new particles with precision measurements
of its own.  For example, the ILC measurements of $\tan\beta$ for Higgs bosons or $\cos\theta_t$ for 
supersymmetric partners of top quarks are unambiguously interpreted at the ILC 
and lead to model-independent measurements.  In this way, the ILC will enable sharper
and more informative analysis of LHC data in addition to the potential for discoveries of its own.
The precision on some such measurements of new particles are listed in Table \ref{tab:NewSummary}

\begin{table}[h]    %[p] 
%\begin{center}
\tbl{Some examples of the precision of measurements of new particles 
that can be achieved by the ILC.}
{\begin{tabular}{lccl}
Topic          &   Parameter & Accuracy  $\Delta X/X$   
 \\  \hline \hline
$ H^0$, $A^0$     &       $ m_H$, $m_A$  &   1.5\%   &  \\ 
                           &   $\tan\beta$          &    20\%    &
                           \\ 
 $ \widetilde \chi^+       $            &
 $  m(\widetilde\chi^+)   $   &    1\%     &  \\
             &
 $  m(\widetilde\chi^0)   $   &    1\%     &  \\
 $ \widetilde t      $            &
 $  m(\widetilde t)   $   &    1\%     &  \\
 &   $   \cos\theta_t    $     &     0.4\%   \\ 
\end{tabular}
\label{tab:NewSummary}}
%\end{center}
\end{table}

As the examples described above illustrate, and the extensive discussion in the ILC Technical Design Report
demonstrate,\cite{tdr-2} the ILC addresses the most important problems of particle physics with a precision
that will significant advance understanding of the underlying physics.  It will provide unprecedented
measurements of the Higgs boson, the W boson, the top quark, as well as other new particles that might
be discovered by the LHC or the ILC.

%% file: collider.tex
%\documentclass{ws-ijmpa}
%\usepackage[super,compress]{cite}
%\usepackage{graphicx}
%\renewcommand{\thefootnote}{\arabic{footnote}}
%\usepackage{textcomp}
%\usepackage{gensymb}
%\begin{document}
%\setcounter{section}{1}

%\section{Physics Purpose and Goals}

%Particle physicists have successfully established a Standard Model of the fundamental particles and their interactions; it has passed numerous tests by many experiments. The Standard Model includes an explanation for the origin of electroweak symmetry breaking through the Higgs mechanism with a Standard Model Higgs boson. A candidate for the Standard Model boson has been discovered at the LHC with properties expected by the theory. But many details of this particle's proper-ties, and the presence or absence of %other anticipated new particles, remain to be discovered to complete the theoretical picture.

%etc.

%\section{The Collider}

The next particle collider should have an energy range for colliding electrons and positrons that complements the physics explored by the Large Hadron Collider at CERN. This implies developing a machine of 200 - 500 GeV that is extendable to 1 TeV or higher. In this report, we describe a collider designed to cover this energy range that is based on Superconducting RF technology.\cite{maury}  An alternate technology would be required to reach much beyond 1 TeV. Two possibilities under study, but still requiring years of R\&D, are an electron-positron linear collider, based on a two-beam concept (CLIC),\cite{CLIC, CLIC-1, CLIC-2} and a muon collider, based on capturing pion decay muons and cooling them to fill counter rotating muon storage rings.\cite{muons}

Over the past 50 years, three generations of particle accelerators have provided major discoveries and advances in particle physics by the complementarity of colliding protons on protons, a broad-band discovery device, with colliding electrons on positrons, as a precision probe. Through research programs using these two approaches, we have made a plethora of discoveries about the basic constituents of matter and fundamental symmetries in nature. 

The International Linear Collider is an electron-positron collider specifically designed to be complementary to the Large Hadron Collider (LHC) at CERN. The ILC will make possible precision studies of the underlying physics of the Higgs, as well as provide a different way of discovering the physics of this new energy regime.
  
Developing such a companion electron-positron machine presents an extraordinary challenge. Electrons and positrons are $\sim$2,000 times lighter than protons and, consequently, they radiate away much of their energy when bent around a circular orbit in a traditional collider at high energies. In fact, the LEP accelerator was limited by such radiation and a new approach is needed to reach the higher energy scale being explored by the LHC. 

This goal has stimulated the development of linear colliders, consisting of two linear accelerators, one for electrons and the other for positrons colliding their parti­cle beams with each other. The linear collider scheme solves the radiation problem, but introduces a whole set of new problems that come with a single-pass machine. In contrast to having counter-rotating circular colliders where the particles go around the machine multiple times, the beams in a linear collider pass through each accel­erating element only once. Therefore, these accelerating elements must be very efficient at transferring energy to the particles. In addition, at the collision point the beams also cross each other only once so that very dense particle bunches must be produced to achieve the needed collision probability.

\begin{figure}[h]
\centerline{\includegraphics[width=11.8cm]{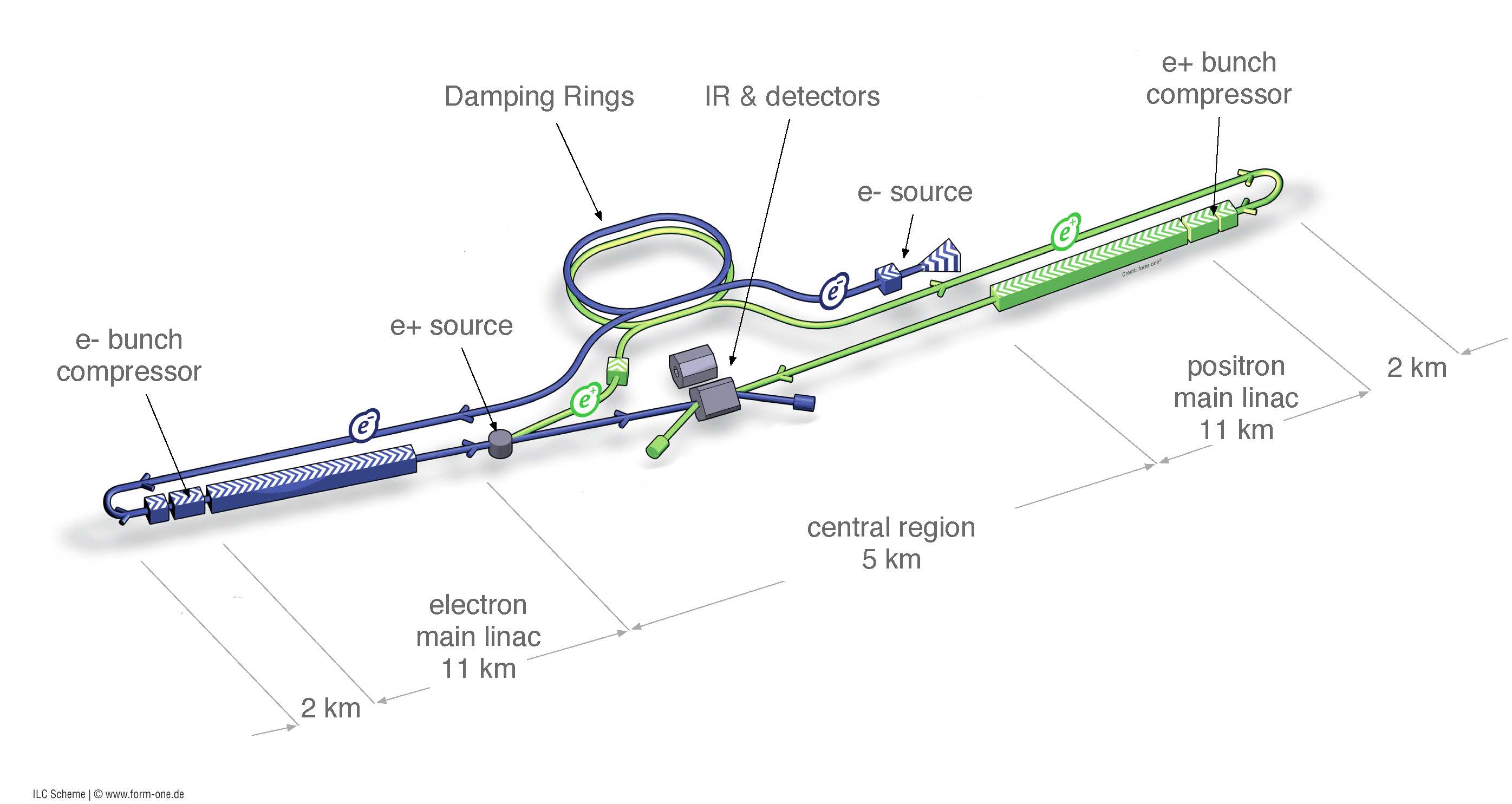}}
\caption{Overall layout of the ILC. \label{f1}}
\end{figure}

Ambitious R\&D programs were carried out towards a linear collider in the 1990s, especially at SLAC and KEK on room temperature technologies and at DESY on superconducting technologies, and they successfully demonstrated the viability of both technologies. The International Committee for Future Accelerators (ICFA) then took on the role of defining the physics goals and choosing the technology (superconducting radio-frequency cavities) to be the basis for the design of the linear collider. ICFA also formed the Global Design Effort (GDE) that has carried out the enabling R\&D program and accelerator design reported here.
	
The International Linear Collider, described in this report, is a linear electron-positron collider based on 1.3 GHz superconducting radio-frequency (SCRF) accelerating technology. It is designed for variable center-of-mass energy 200-500 GeV (extendable to 1 TeV) with high luminosity.  

\subsection{General Features}

The International Linear Collider shown schematically in Fig.~\ref{f1} has energy adjustable from 200 to 500 GeV, and is upgradeable to 1 TeV with the addition of more linac units.  Although the main linac is by far the largest, most complex and challenging part of the machine, there are considerable challenges involved in producing electron and positron beams that satisfy the requirements for injection into the two linacs.

The injection subsystems include: a polarized electron source that is based on a photocathode DC gun; a polarized positron source in which positrons are obtained from electron-positron pairs by converting high-energy photons produced by passing the high-energy main electron beam through an undulator; 5 GeV electron and positron damping rings, each having a circumference of 3.2 km, housed in a common tunnel; beam transport systems from the damping rings to the main linacs, followed by a two-stage bunch compressor system before injection.  The two 11 km main linacs utilize 1.3 GHz SCRF cavities operating at an average gradient of 31.5 MV/m, and have a pulse length of 1.6 msec. After being accelerated in the linacs the electron and positron beams are ejected into two opposite-facing beam-delivery systems, each 2.2 km long and meeting at a single collision point that has a 14 mrad crossing angle.  The interaction point is shared by two detectors, arranged in a `push-pull' configuration.

\begin{figure}[h]
\centerline{\includegraphics[width=8.8cm]{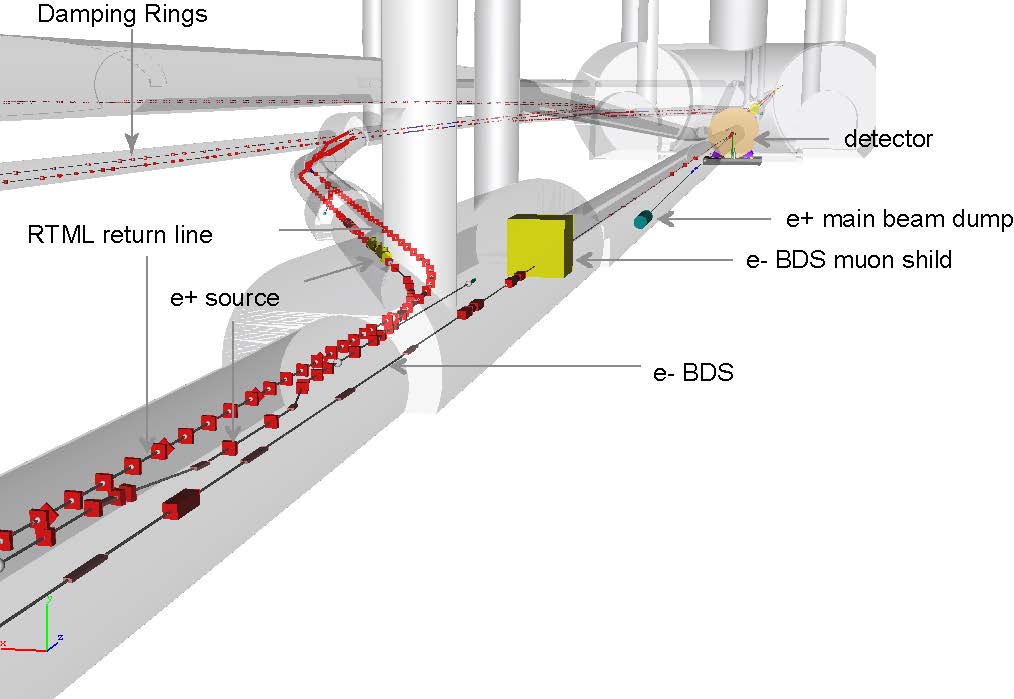}}
\caption{The complex ILC central region contains many of the key ILC subsystems. \label{f2}}
\end{figure}

The total length of the ILC 500 GeV complex is $\sim31$ km long. The central 5 km region
shown in Fig.~\ref{f2}
contains many of the key accelerator subsystems.  The electron source, positron source (including an independent low-powered auxiliary source), and the electron and positron damping rings are all centrally located around the interaction region (IR). The damping-ring complex is displaced laterally to avoid interference with the detector hall. The electron and positron sources themselves are housed in the same (main accelerator) tunnels as the beam-delivery systems, which reduces the overall cost and size of the central-region underground construction.

\subsection{Machine Parameters}

A general survey of ILC physics was carried out by a subcommittee of ICFA that resulted in a set of top level physics goals \cite{ICFA} that the GDE used as the physics  requirements to flow down into the basic ILC machine technical requirements.  The top level physics goals from the study include:

\begin{itemlist}
 \item Luminosity $\rightarrow \int Ldt = 500$ fb$^{-1}$ in 4 years
 \item E$_{cm}$ adjustable from 200--500 GeV
 \item Ability to scan between 200 and 500 GeV
 \item Energy stability and precision below 0.1\%
 \item Electron polarization of at least 80\%
\end{itemlist}

In addition, the ICFA report identified several high-priority options that should be included in the design: the machine must be upgradeable to 1 TeV and also include the possibility of positron polarization is desirable.

The ILC design reported here achieves these design performance physics goals and has a set of top-level technical design parameters that are summarized in Table~\ref{table1}.

\begin{table}[h]
\tbl{The top level operating parameters at 500 GeV.}
{\begin{tabular}{@{}lcl@{}} \toprule
Max. Center-of-mass energy & 500 & GeV  \\
\colrule
Peak Luminosity & $\sim{2}\times10^{34}$ & $1/cm^2s$ \\
Beam Current & $9.0$ & mA \\
Repetition rate & $5$ & Hz \\
Average accelerating gradient & $31.5$ & MV/m \\
Beam pulse length & $0.95$ & ms \\
Total Site Length & $31$ & km \\
Total AC Power Consumption & $\sim{230}$ & MW \\
\botrule
\end{tabular} \label{table1}}
\end{table}

Following the discussion above, the ILC is designed to have an operational range from 200 to 1000 GeV, where the design parameters have been optimized for cost and risk, as well as the physics performance.  The design has relatively conservative operating points that take into account constraints imposed by the various accelerator sub-systems, such as various instability thresholds (most notably the electron cloud in the positron ring), realistic rise-times for the injection and extraction kickers, and the desire to minimize the circumference of the rings.  

Positron production in the undulator-based source is significantly degraded for electron beam energy below 150 GeV.
Therefore, for center-of-mass operation below 300 GeV the repetition rate is increased to 10 Hz, with
alternate electron pulses used for positron production (150 GeV) and collisions ($>$ 150 GeV).

The maximum length of the beam pulse is constrained to $\sim1.6$ msec, which has been routinely achieved in the available 1.3 GHz 10 MW multi-beam klystrons and modulators.  The beam current is further constrained by the need to minimize the number of klystrons (peak power) and higher-order modes (cryogenic load and beam dynamics). The repetition rate of the machine is a cost driver that limits the dynamic cryogenic load (refrigeration).
 
The electron and positron sources constrain the achievable beam current and the total charge: for the laser-driven photocathode polarized electron source, these limits are set by the laser, while for the undulator-based positron source, the limits are set by the power deposition in the photon target. 

The beam pulse length is further constrained for both electrons and positrons by the achievable performance of the warm RF capture sections. Finally, at the interaction point, the single-bunch parameters are limited by the strong beam-beam effects and the detector requirements on both the beam-beam backgrounds and beam stability.

\subsection{SCRF Main Linac}

The ILC design is based on high-gradient superconducting RF technology for the main linac.  An intensive R\&D program has been carried out to achieve the high gradient performance. In addition, this R\&D has been combined with a concerted effort to develop a practical scheme for building and assembling a high performance main linac from in-kind components that will be built and contributed from around the world.

\begin{figure}[h]
\centerline{\includegraphics[width=10.8cm]{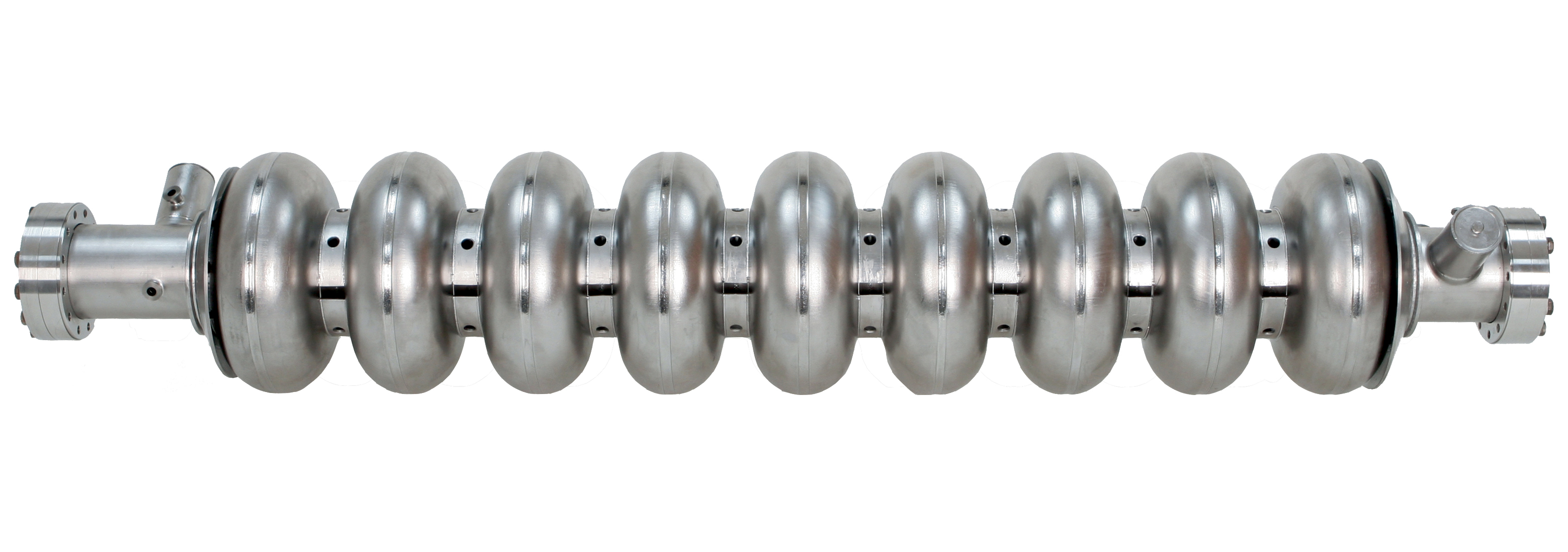}}
\caption{1 m long, nine-cell 1.3 GHz superconducting niobium cavity. \label{f3}}
\end{figure}

The baseline cavity (Fig.~\ref{f3})
used for the ILC design is the TESLA 9-cell superconducting cavity that has been developed over the past 15 years and has achieved the highest gradients to date for multi-cell cavities.  Numerous cavities of this style have been manufactured by different vendors and processed at various laboratories around the world. Each cavity consists of nine accelerating cells and two end-group sections. One end group has a port for coupling RF power from the power source into the structure, and the other end has a port for a field-sampling probe used to determine and control the accelerating gradient. Each end group also has a resonant higher-order-mode (HOM) coupler structure with a probe port that is connected to a small electric-field antenna for extracting HOM power and for diagnostics.

Achieving the required high gradient performance has been possible through a directed SCRF R\&D program to identify the technical causes for gradient limitations and developing ways to eliminate these problems. The application of these remediation methods has enabled defining a baseline process for consistent production of 35 MV/m cavities.

\begin{figure}[h]
\centerline{\includegraphics[width=10.8cm]{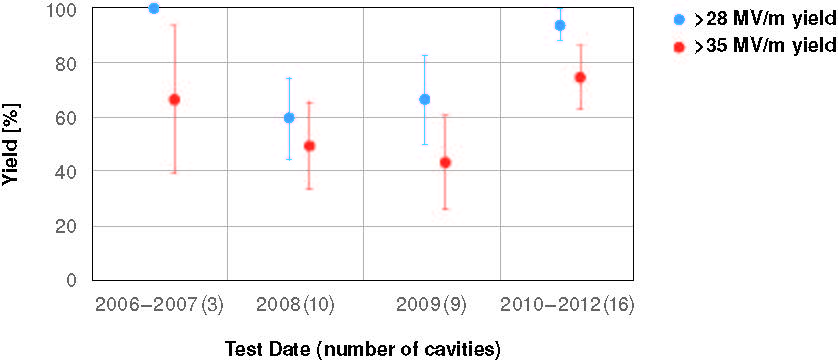}}
\caption{Cavity yield for two gradient thresholds vs. time for two passes using a standard prescribed treatment.  The numbers in parentheses are the cavity sample size. \label{f4}}
\end{figure}

The high-gradient R\&D goal was established to demonstrate a production yield of 90\% worldwide.  This is a practical cost effective yield for industrially fabricated cavities.  With limited statistics, our tests (Fig.~\ref{f4}) have achieved a yield of 94\% for cavity production above 28 MV/m having an average gradient of 37.1 MV/m.  This result is consistent with the gradient-spread specification of  $\pm 20\%$.  Much larger statistics on yields will soon be obtained from the European XFEL cavity production program, for which at least the first-pass processing is very similar to that of the ILC.

For the 500 GeV ILC, each main linac accelerates the beam from 15 GeV (after acceleration in the upstream bunch compressors) to a maximum energy of 250 GeV. Beam acceleration is provided by approximately 7,400 one meter-long superconducting nine-cell niobium cavities operating at 2 degrees K and assembled into  $\sim$850 cryomodules. The average accelerating gradient of the cavities is 31.5 MV/m at 500 GeV center-of-mass beam energy and with a corresponding quality factor $Q_{0}>10^{10}$.

The main linacs follow the Earth's average curvature.  In general this enables siting within a favorable geological layer, and also simplify the cryogenics by keeping a constant pressure.   Finally, civil construction simplifies and is possibly less expensive.
 
A random cavity-to-cavity gradient spread of $\pm 20\%$ is accommodated to take into account the expected mass production variations in the maximum achievable gradient. The choice of these key parameters is the result of over a decade of extensive R\&D. The GDE recognized the need to establish expertise in this technology in all three regions of the world and established the high-gradient program as its highest priority during the Technical Design Phase. As a result, extensive worldwide experience both in the labs and in industry now gives high confidence that these requirements can be routinely achieved.

The optimal matched $\mathit{Q}_L$ is $\sim 5.4\times10^6$ for operation of the linac having average gradient of 31.5 MV/m and nominal beam current of 5.8 mA. This corresponds to a cavity fill time of $925 \hfill\mu{sec}$. Added to the nominal beam pulse width of $727 \hfill\mu{sec}$, this gives a total RF pulse length of 1.65 msec

\begin{figure}[h]
\centerline{\includegraphics[width=10.8cm]{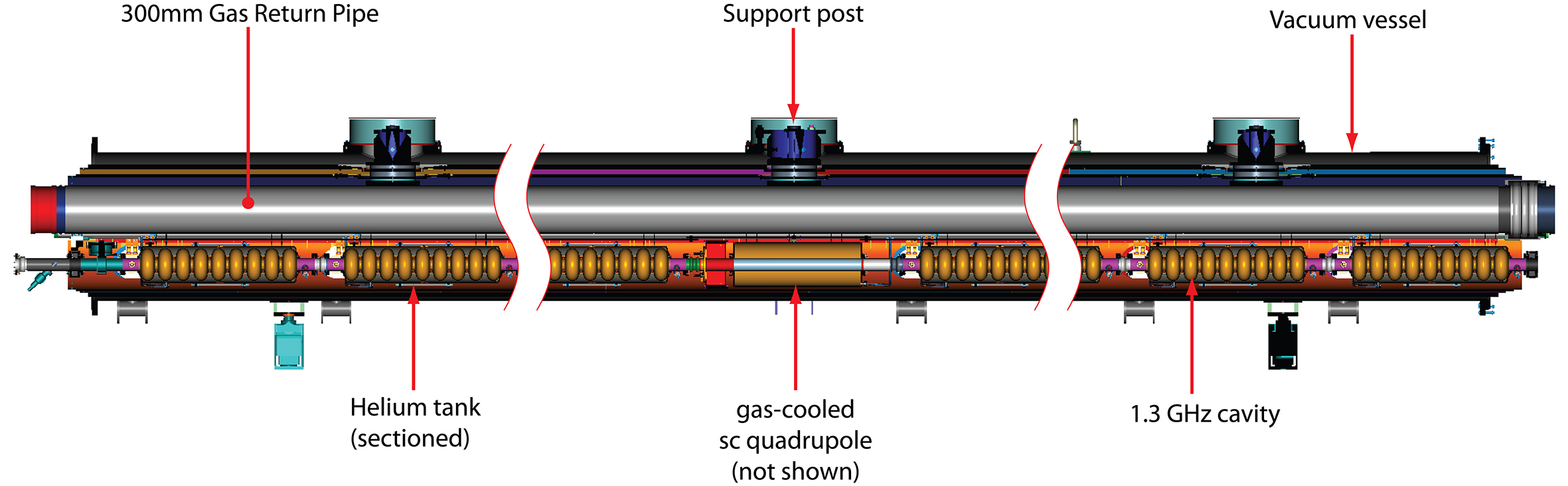}}
\caption{Longitudinal cross section of a Type B ILC cryomodule. \label{f5}}
\end{figure}

The cavity package illustrated in Fig.~\ref{f5}
includes the cavity mechanical tuner, which is integrated into the titanium helium vessel of the cavity, and an adjustable high-power coupler. In addition to a slow mechanical tuner (used for tuning and initial slow drift compensation), a fast piezo-driven tuner is also included to compensate dynamically for the mechanical deformation of the cavity caused by the RF pulse, known as ``Lorentz-force detuning."

The cryomodules (Fig.~\ref{f5}) that make up the Main Linacs are 12.65 m long. There are two types: a Type A module with nine 1.3 GHz nine-cell cavities and Type B with eight nine-cell cavities and one superconducting quadrupole package located at the center of the module.

The ILC cryomodule design is a modification of those developed and used at DESY in the FLASH accelerator as well as the 100 cryomodules currently being produced by industry for the European X-Ray FEL (XFEL), also based at DESY. A 300 mm-diameter helium-gas return pipe serves as the primary support for the nine cavities and other beamline components in the Type A module. For Type B, the central cavity package is replaced by a superconducting quadrupole package that includes the quadrupole itself, a cavity BPM, and superconducting horizontal- and vertical-corrector dipole magnets. The quadrupoles establish the magnetic lattice for the Main Linac, which is a weak-focusing FODO optics with an average beta function of $\sim 80$ m. Every cryomodule also contains a 300 mm-long assembly that removes energy from beam-induced higher-order modes above the cavity cut-off frequency through the 40-80 K cooling system.

To operate the cavities at 2 K, they are immersed in a saturated He-II bath. Shields are cooled with Helium gas to intercept thermal radiation and provide a heat sink for conduction at 5-8 K and at 40-80 K. Each cryomodule has an estimated 2 K static cryogenic heat load of 1.3 W while the 2 K dynamic heat load is approximately 9.8 W. Liquid helium for the main linacs and the bunch compressor RF is supplied from a total of 10-12 large cryogenic plants, each of which has an installed equivalent cooling power of $\sim20$ kW at 4.5 K. The plants are located in pairs separated by 5 km along the linacs, with each plant cooling $\sim2.5$ km of continuous linac. 

The RF power is provided by 10 MW multi-beam klystrons (MBK) each driven by a 120 kV Marx modulator. The 10 MW MBK has achieved the ILC specifications and is now a well-established technology with several vendors worldwide. The 120 kV Marx-modulator prototypes have achieved the required specifications and are now undergoing design optimization for transfer to industrial vendors.

Two alternative methods of transporting the RF microwave power to the accelerating structures have been presented in the baseline, one more practical for a mountainous site and the other for a deep underground flat site. 

A Distributed Klystron Scheme (DKS) is proposed for a mountainous site, where access from the surface is limited and distant.  In this case, each klystron drives 39 cavities and the klystrons and modulators are distributed along the entire length of the SCRF linacs.  The are underground in the same tunnel, but shielded from the accelerator itself, which enables beam-on personnel access to make repairs.  

A novel Klystron Cluster Scheme (KCS), where all the klystrons are located in `clusters' in surface buildings located periodically along the linacs is proposed for a deep underground flat site. The power from each cluster of 19 klystrons ($\sim 190$ MW) is combined into an over-moded waveguide and the power is transported to the underground the tunnel to service approximately a 1 km section of linac. A Coaxial Tap-Off extracts $\sim 6.7$ MW of power to a local power-distribution system feeding three cryomodules containing 26 cavities.

The KCS has the feature that most of the heat load is on the surface, where it can be more cost-effectively removed and at the same time reducing the required underground volume. However, it requires additional surface buildings and shafts (one every 2 km of linac) and there are additional losses in the long waveguide distribution systems. Finally, this scheme still requires significant R\&D compared to the mature and tested DKS system. Nonetheless, the estimated cost savings associated with KCS make it a more attractive solution for flat deep underground sites.  

For both KCS and DKS, the in-tunnel power-distribution system to the cavities is essentially identical. A key common feature and requirement is the capability to vary the phase and forward power to each cavity remotely, matching the required $\pm20\%$ cavity gradient spread.  This enables maximizing the accelerating gradient for the entire linac.

\subsection{Other Accelerator Subsystems}

\subsubsection{Electron Source}

\begin{figure}[h]
\centerline{\includegraphics[width=10.8cm]{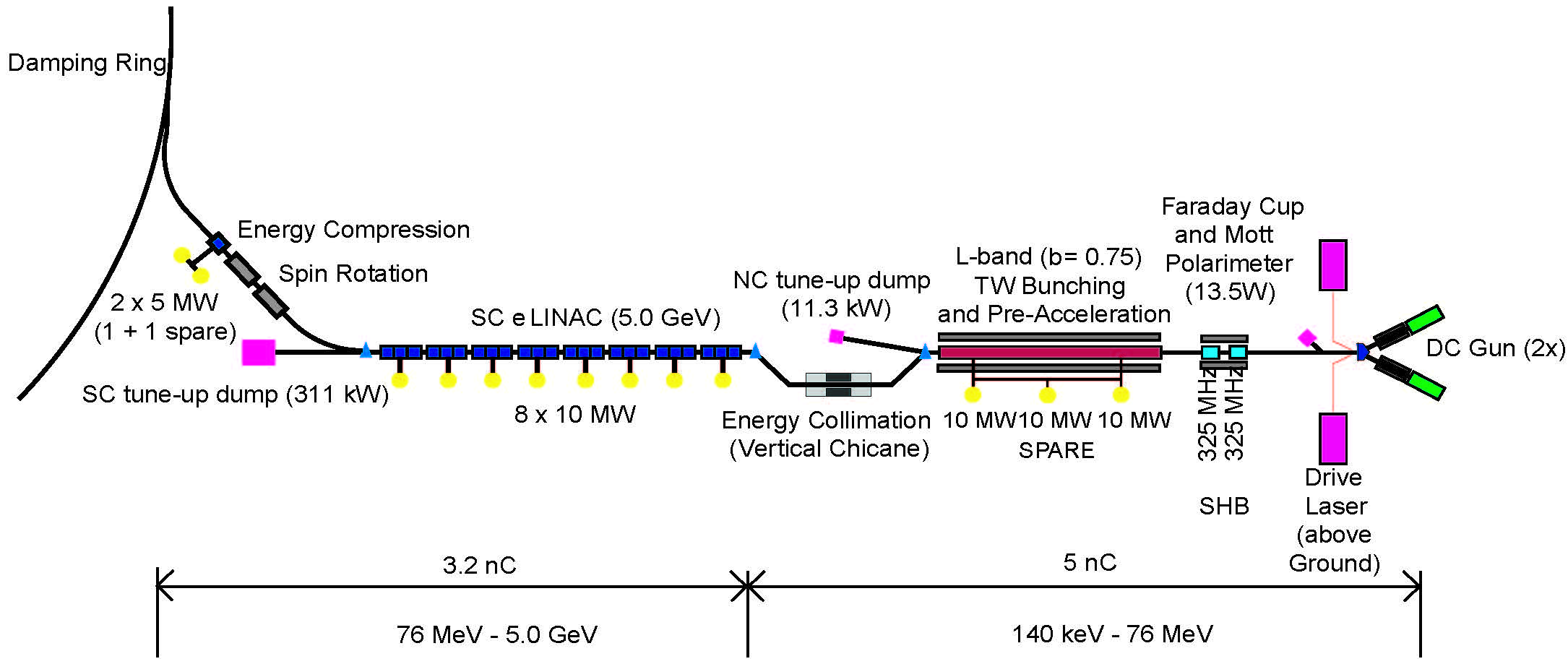}}
\caption{Schematic view of the Polarized Electron Source.
\label{f6}}
\end{figure}

The electron source (Fig.~\ref{f6}) shares the central region accelerator tunnel with the positron Beam Delivery System.  The electron beam is produced by a laser illuminating a strained GaAs photocathode in a DC gun, providing the necessary bunch train that is transported to the damping ring. It has 1312 bunches of $2.0\times10^{10}$ electrons at 5 Hz with polarization greater than 80\%. Two independent laser and gun systems provide redundancy.  Normal-conducting structures are used for bunching and pre-acceleration to 76 MeV, after which the beam is accelerated to 5 GeV in a superconducting linac. Before injection into the damping ring, superconducting solenoids rotate the spin vector into the vertical, and a separate Type-A superconducting RF cryomodule is used for energy compression. 

The primary challenge for the ILC electron source is the long bunch train, which requires a laser system beyond the present state of the art on accelerators, and normal conducting structures capable of the high RF power. R\&D prototypes have been tested to demonstrate the feasibility of both systems.

\subsubsection{Positron Source}

\begin{figure}[h]
\centerline{\includegraphics[width=10.8cm]{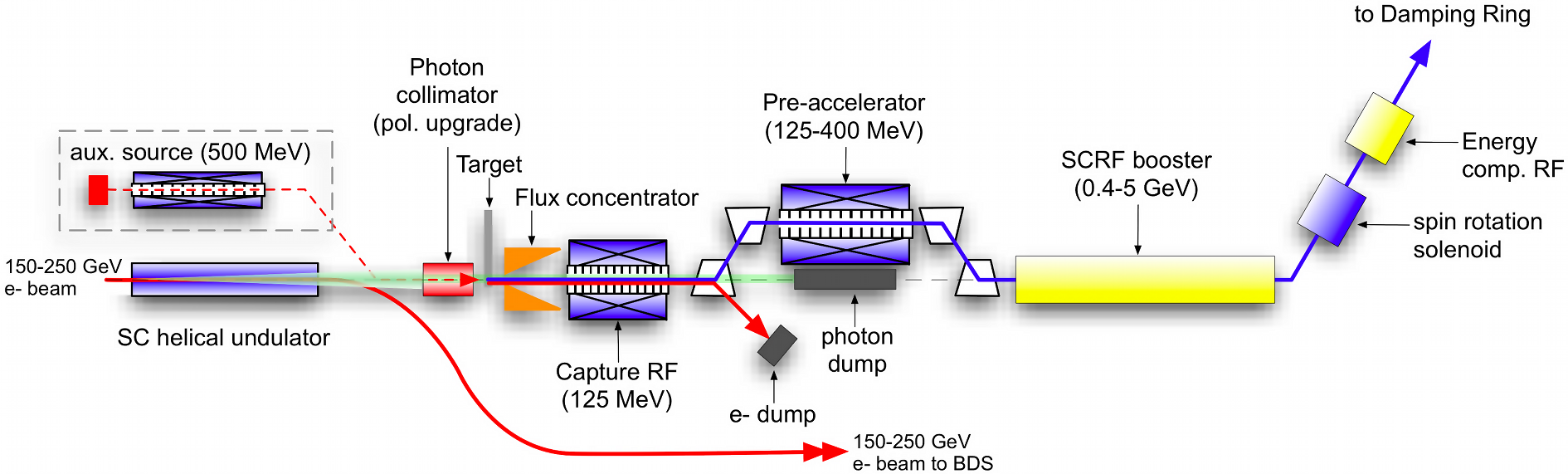}}
\caption{Schematic view of the Positron Source.
\label{f7}}
\end{figure}

The ILC positron source (Fig.~\ref{f7}) makes the positron beam beginning from the electron main linac beam, passing it through a long helical undulator to generate multi-MeV photons that strike a thin metal target, creating positrons from the  electromagnetic showers. Positrons are separated from the other shower constituents and unused photons, focused, captured into a beam, accelerated and transported to the Damping Rings.

The baseline design is to develop a beam with 30\% polarization. There are spin rotators before injection into the damping rings that preserve the polarization, and there is sufficient beamline space to allow for an upgrade to  $\sim 60\%$ polarization in the future.

Some key features of the positron source are:

\begin{itemlist}
 \item The generation of a high-power multi-MeV photon beam for producing positrons. This requires suitably short-period, high-K-value helical undulators; 
 \item The production of positron bunches in a metal target that can handle the beam power and radioactive environment resulting from the production process. This requires a high-power target system; 
 \item The capture, acceleration and transport of the positron bunch to the Damping Rings with minimal beam loss. This requires high-gradient normal-conducting RF and special magnets to capture the positrons efficiently and to transport the resulting large transverse emittance positron beams.  
\end{itemlist}

The photons from the undulator are directed onto a rotating 0.4 radiation-length Ti-alloy target $\sim$500 m downstream, producing a beam of electron-positron pairs. This beam is then matched using an optical-matching device (a pulsed flux concentrator) into a normal conducting (NC) L-band RF and solenoidal-focusing capture system and accelerated to 125 MeV. The electrons and remaining photons are separated from the positrons and dumped. The positrons are accelerated to 400 MeV in a NC L-band linac with solenoidal focusing. Similar to the electron beam, the positron beam is then accelerated to 5 GeV in a superconducting linac which uses modified Main Linac cryomodules, the spin is rotated to the vertical, and the energy spread compressed before injection into the positron damping ring.

The target and capture sections are high-radiation areas which will require shielding and remote handling facilities.  The baseline design provides a polarization of 30\%. There is space left to insert a future additional $\sim220$ m undulator for an upgrade to 60\% polarization, which would also require a photon collimator upstream of the target.

A low-intensity auxiliary positron source supports commissioning and tuning of the positron and downstream systems when the high-energy electron beam is not available. This is accomplished with a conventional positron source that uses a 500 MeV room temperature linac to provide an electron beam that is directed onto the photon target, providing a few percent of the design positron current.

In order to accommodate 10 Hz operation required for center-of-mass energies below 300 GeV, a separate pulsed extraction line is employed after the undulator to transport the 150 GeV electron pulse for positron-production to the high-powered tune-up dump.

\subsubsection{Damping Rings - Description}

The electron and positron input beams to the ILC damping rings have large transverse and longitudinal emittances.  The role of the damping rings is to reduce these emittances to that required to transport down the main linac and obtain the desired small beam spot needed to achieve design luminosity. The specification for the extracted normalized vertical emittance is 20 nm and that represents a reduction of five orders of magnitude for the positron bunch. This reduction must be achieved within the 200 msec between machine pulses (100 msec for 10-Hz mode). 

The $\sim 1$ msec beam pulse must be compressed on injection by roughly a factor of 90, in order to fit into the ring circumference of 3.2 km and a corresponding decompression is required on extraction. For the baseline parameters, the bunch spacing within trains is approximately 8 ns and this determines the rise and fall time of the injection and extraction kicker systems. (For a luminosity upgrade this number reduces to $\sim 4$ ns.) Individual bunch injection and extraction is accomplished in the horizontal plane using a total of 42 fast kickers switching 10 kV pulses with rise/fall times of $\sim 3$ ns.

\begin{figure}[h]
\centerline{\includegraphics[width=8.8cm]{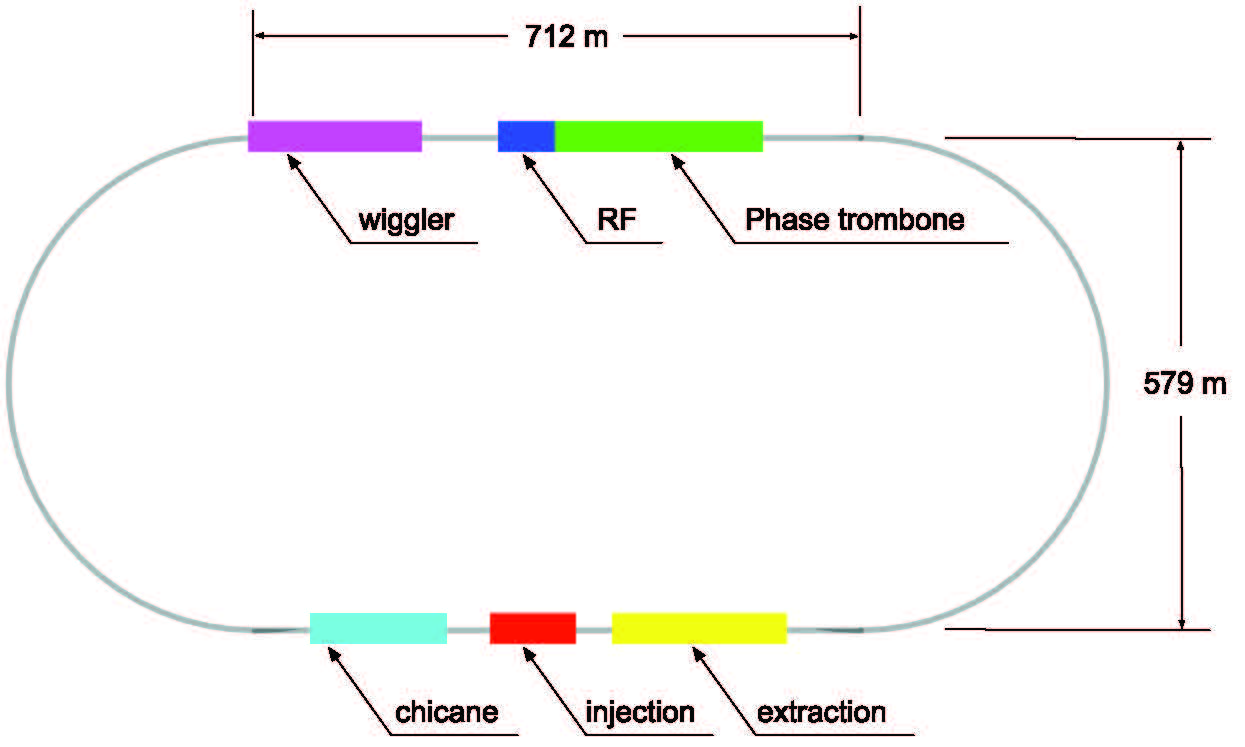}}
\caption{The damping ring layout.  The circumference is 3238.7 m and the length straight sections are 710.2 m.
\label{f8}}
\end{figure}

The baseline design as shown in Fig.~\ref{f8} has one electron and one positron ring that operate at  beam energies of 5 GeV. Both rings are housed in a single tunnel with one ring positioned directly above the other. The damping ring complex is located in the central region, horizontally offset from the interaction region by approximately 100 m, in order to avoid the detector hall. 

The damping-ring lattice follows a race-track design. The two arc sections are constructed from 75 Theoretical Minimum Emittance (TME) cells. One of the two 712 m-long straight sections accommodates the RF cavities, damping wigglers, and a variable path length to accommodate changes in phase (phase trombone), while the other contains the injection and extraction systems, and a circumference-adjustment chicane. Damping is accomplished by approximately 113 m of superferric wigglers (54 units $\times$ 2.1 m) in each damping ring. The wigglers operate at 4.5 K, with a peak-field requirement of 2.16 T.

The superconducting RF system is operated in continuous-wave (CW) mode at 650 MHz, and provides a maximum of 20 MV for each ring, required for the positron ring in 10 Hz mode (nominal 5 Hz operation requires 14MV for both electron and positron). The frequency is chosen to be half the linac RF frequency to maximize the flexibility for different bunch patterns. The single-cell cavities operate at 4.5 K and are housed in twelve 3.5 m-long cryomodules. The RF section of the lattice can accommodate up to 16 cavities, of which 12 are assumed to be installed for the baseline.

The momentum compaction of the lattice is relatively large, which helps to maintain single-bunch stability, but requires a relatively high RF voltage to achieve the design RMS bunch length (6 mm). The dynamic aperture of the lattice is sufficient to allow the large-emittance injected beam to be captured with minimal loss.

Mitigation of the fast ion instability in the electron damping ring is achieved by limiting the pressure in the ring to below 1 nTorr and by the use of short gaps in the ring fill pattern and a fast transverse feedback system, similar to those used in B-factories.

\subsubsection{Damping Rings - Electron Cloud Mitigation}

A particular challenge for the ILC design is the performance of the low emittance positron damping ring.  In particular, electrons emitted from the vacuum-pipe walls by synchrotron-radiation photons create a cloud of electrons that can defocus the positron beam, increases the beam emittance. A program to understand electron cloud induced emittance growth and beam instabilities was undertaken, including tests of mitigation techniques to reduce the effects to maintain the design beam optical parameters.  

Simulation studies indicated that the effects are large enough in the positron damping rings that mitigation is required. Therefore, an ambitious R\&D program has been carried out at the Cornell Electron-Positron Storage Ring Test Accelerator (CesrTA) to study the dynamics of electron cloud effects and the effectiveness of various mitigation techniques. Simulation studies and tools have been developed and baselined, mitigation techniques tested, and results from other laboratories and earlier studies have been used to develop recommended mitigation schemes.

Two main goals at CesrTA were to characterize the build-up of the electron cloud in each of the key magnetic field regions, particularly in the dipoles and wigglers, and to study the most effective methods of mitigation in each region. This required the design and installation of detectors to study the local build-up and a simulation program to characterize the results. The second goal was to understand the impact of the electron cloud on ultra-low-emittance beams similar to those for the ILC damping rings.  

Benchmarking electron-cloud instability and emittance-growth simulations with this modified CesrTA configuration has enabled making reliable estimates for the ILC.  CESR had to be reconfigured as a low-emittance damping ring and instrumented to monitor and make the measurements with the necessary beam instrumentation.

Converting CESR into a low emittance damping-ring was accomplished by relocating six of the twelve damping wigglers to straight sections to enable ultra-low-emittance CesrTA operations, upgrading the beam instrumentation to enable high-resolution beam-position monitoring, and adding vacuum-system diagnostics for characterizing local electron-cloud growth.  At 2 GeV, 90\% of the synchrotron-radiation power is provided by the twelve damping wigglers and a horizontal design emittance of 2.6 nm rad was achieved. A vertical emittance target of less than 20 pm rad (ten times the ILC damping-ring vertical-emittance target) was specified.  CESR allows operation between 1.8 and 5.3 GeV with both positron and electron beams and this flexibility in energies, bunch spacings and intensities has enabled systematic studies of primary photoelectron and secondary electron contributions to electron-cloud build-up in the vacuum chambers.

\begin{figure}[h]
\centerline{\includegraphics[width=8.8cm]{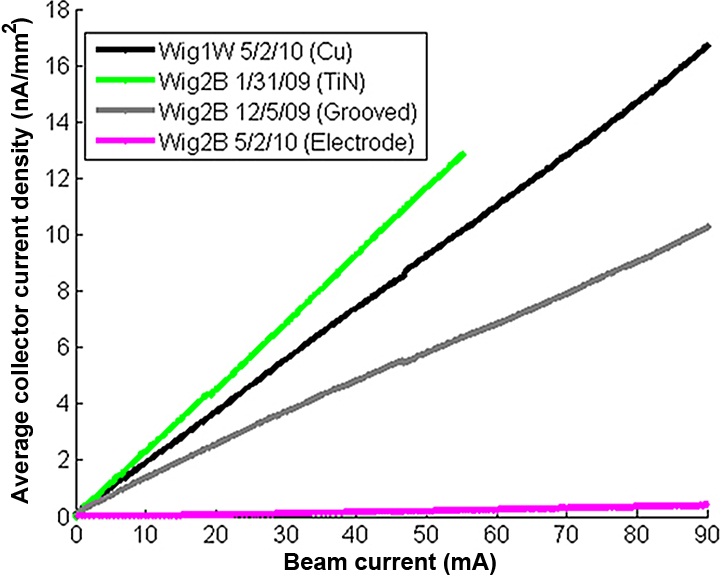}}
\caption{A comparison of the measured retarding-field analyzer current in a wiggler versus beam current with a 20-bunch positron train for a bare copper surface, a titanium-nitride-coated copper surface, a grooved copper surface and a clearing electrode.
\label{f9}}
\end{figure}

Three different mitigation techniques have been studied:  1) coating the vacuum pipe surface with a titanium-nitride-coated copper surface, 2) grooving the copper surface, and 3) employing clearing electrodes.  A comparison of the effectiveness vs beam current is shown in Fig. \ref{f9}. The data indicate that the best cloud suppression in the wiggler region is obtained with clearing electrodes.

\begin{table}[h!]
   \tbl{Baseline mitigation scheme for electron cloud effects in the ILC TDR damping rings.}{
\label{tab:ECWG_Recommendations}
   \begin{tabular}{l p{1.0in} ll}
\toprule
Field  & \multicolumn{2}{c}{Mitigation Recommendation}  & \multicolumn{1}{c}{Alternatives for}      \\
{Region} & \multicolumn{1}{c}{\textit{Primary}} & \multicolumn{1}{c}{\textit{Secondary}} & \multicolumn{1}{c}{Further Study} \\
\colrule
Drift* & TiN Coating                                & Solenoid Windings                            & NEG Coating \\\noalign{\smallskip}
Dipole & \multicolumn{1}{p{0.9in}}{Triangular Grooves with TiN Coating}        & \multicolumn{1}{p{1.25in}}{Antechambers for synchrotron radiation %
   power loads and photoelectron control}        & \multicolumn{1}{p{1.3in}}{R\&D into the %
                            use of clearing electrodes} \\
Quad* & TiN Coating                            &                                               & \multicolumn{1}{p{1.3in}}{R\&D into the use of clearing electrodes or grooves with TiN coating}                        \\\noalign{\smallskip}
   {Wiggler} & \multicolumn{1}{p{0.9in}}{Clearing Electrode}                       & \multicolumn{1}{p{1.25in}}{Antechambers for synchrotron radiation power loads and photoelectron control}        & \multicolumn{1}{p{1.3in}}{Grooves with TiN coating}  \\\noalign{\smallskip}
   \botrule
   \multicolumn{4}{p{0.95\textwidth}}{* \footnotesize{Where drift and quadrupole chambers are in arc or wiggler straight regions of the machine, the chambers will
incorporate features of those sections, i.e. antechambers for power loads and photoelectron control.}
}
   \end{tabular}}
\end{table}

Following studies of mitigation effectiveness, a baseline mitigation scheme for the ILC TDR has been 
adopted (Table~\ref{tab:ECWG_Recommendations}). 
It should be noted that the choices of mitigation methods in the table are nearly identical to the choices that have been made for the construction of the SuperKEKB vacuum system. Thus operation of the SuperKEKB positron ring will serve as a crucial performance test that will further improve understanding of the anticipated performance of the ILC damping rings.

\subsubsection{Ring to Main Linac (RTML)}

The electron and positron Ring to Main Linac (RTML) systems are the longest continuous beamlines in the ILC. The layout of the RTML is identical for both electrons and positrons. The RTML consists of five subsystems, representing the various functions that it must perform: a $\sim15$ km long 5 GeV transport line; betatron and energy collimation systems; a 180$^\circ$ turn-around, which enables feed-forward beam stabilization; spin rotators to orient the beam polarization to the desired direction; and a two-stage bunch compressor to compress the beam bunch length from several millimeters to a few hundred microns, as required at the IP.

The two-stage bunch compressor includes acceleration from 5 GeV to 15 GeV in order to keep the increase in relative energy spread associated with bunch compression small. The acceleration is provided by sections of SCRF main-linac technology. A primary challenge for the RTML systems is the preservation of the damped emittance extracted from the damping rings; the combination of the long uncompressed bunch from the damping ring and large energy spread (after compression) make the tuning and tolerances particularly demanding. However, tuning techniques developed through detailed simulations have demonstrated acceptable emittance growth.

In addition to the beam-dynamics challenges, acceptable jitter in bunch arrival time at the IP requires an RMS phase jitter of $\sim 0.24^\circ$ between the electron and positron bunch-compressor RF systems. Beam-based feedback systems integrated into the bunch-compressor low-level RF system should be able to limit the phase jitter to this level.

\subsubsection{Beam Delivery System}

Achieving a stable $\sim 10$ nanometer vertical beam spot is the difficult beam requirement to achieve design luminosity.   As discussed earlier, this requires very demanding performance on the input optics, especially the damping rings.   In that regard, we have paid special attention to mitigating the electron cloud effects in the positron damping rings.    In addition to its primary job of accelerating the beams, the main linac must preserve the low emittance beam.  Finally, the ILC Beam Delivery System (BDS) transports the $e^+$ and $e^-$ beams from the exits of the main linacs, focuses them to the very small spot sizes required to reach high luminosity and brings them into collision at the interaction point.   

The remaining beams are sent into beam dumps.  The BDS must perform several other critical functions, including characterizing the incoming (transverse) beam phase-space and matching it into the final focus; removing beam halo from the linac to minimize background in the detectors; and measuring and monitoring the key physics parameters such as energy and polarization before and after the collisions.

There is a single shared collision point with a 14 mrad total crossing angle.  Two detectors alternatively are positioned at the collision point through a `push-pull' configuration.  The 14 mrad geometry provides enough space for separate extraction lines, however crab cavities are required to rotate bunches in the horizontal plane to effectively make head-on collisions. 

\begin{figure}[h]
\centerline{\includegraphics[width=10.8cm]{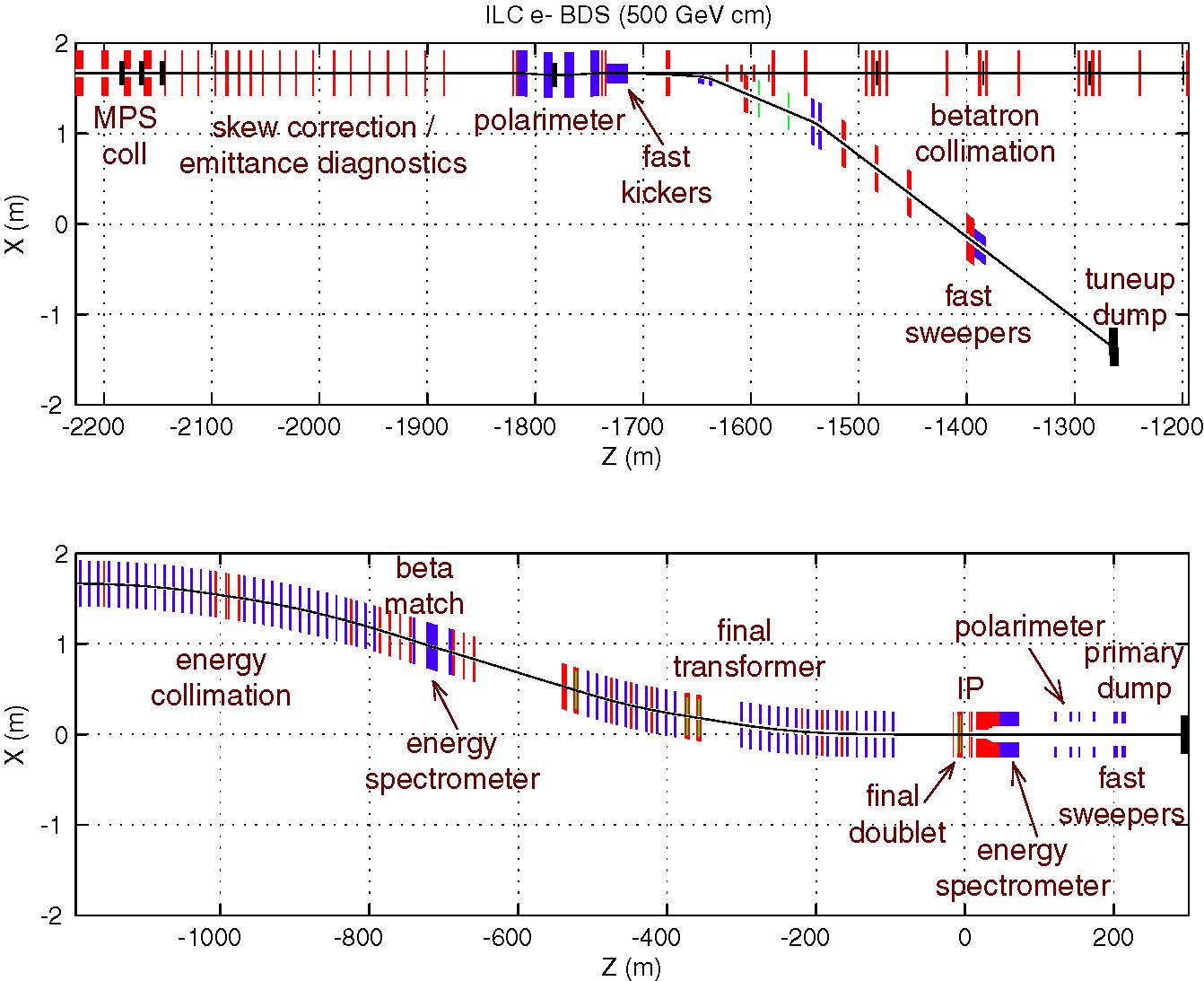}}
\caption{BDS layout showing the horizontal position of the subsystems, beginning from the linac exit with distance measured from the IP.
\label{f10}}
\end{figure}

The BDS minimizes the emittance growth due to synchrotron radiation to less than 10\% and is designed to accommodate a future upgrade to 1 TeV center-of-mass energy, while the 500 GeV baseline lattice uses fewer magnets (predominantly dipoles) for the lower-energy operation.  

The BDS layout shown in Fig.~\ref{f10}
consists of five main subsystems:  first, a section for emittance measurement and matching (correction), trajectory feedback, polarimetry and energy diagnostics; second, a collimation section to remove beam-halo particles and it has magnetized iron shielding to deflect muons generated in the collimation process; third, the final focus (FF) that uses strong compact superconducting quadrupoles to focus the beam at the IP with sextupoles providing local chromaticity correction; the final pair of quadrupoles closest to the IP is integrated into each particle physics detector to facilitate rapid exchanges of the detectors in `push-pull'; fourth, the detectors surround the IP in the interaction region; and fifth, an extraction line that transports the disrupted beam to a high-powered water-cooled dump, containing polarization and energy diagnostics.

The beam-delivery optics demagnifies the beams by a typical factor of a hundred, resulting in very large beta functions and the tightest alignment tolerances in the ILC. In addition, correction of the strong chromaticity and geometric aberrations requires a delicate balance of higher-order optical terms. The tight tolerances on magnet motion (tens of nanometers), makes continuous trajectory correction and the use of fast beam-based feedback systems mandatory. Furthermore, several critical components (e.g. the final-focusing doublet) may well require mechanical stabilization. 

Beam-based alignment and beam phase-space tuning algorithms are necessary to adjust and tune the optical aberrations that would otherwise significantly degrade the luminosity. The ability to tune the beams to the required levels relies extensively on precision remote mechanical adjustment of the magnets and diagnostics with matching precision. Many of the techniques and instrumentation were successfully developed at the Final Focus Test Beam experiment at SLAC and those developments continue at the ATF2 facility at KEK.

The tight tolerance on the relative uncorrelated phase jitter between the electron and positron superconducting crab-cavity systems requires timing precisions to the level of tens of femtoseconds. Although this tolerance is tight, it is comparable to that achieved at modern linac-driven FELs.

Control of machine-generated backgrounds is performed by careful optics control and tuning of an extensive collimation system, as well as by the use of non-linear elements (`tail-folding' octupoles).  Wakefield effects at the small apertures are taken into account in the design of the mechanical collimators.

The main beam dumps use a high-pressure high-velocity water system. Since the dumps will be activated during operation, the dumps are rated for the full upgrade average beam power of 14 MW.  This avoids having to replace highly radioactive dumps during the energy upgrade to 1 TeV.

\subsubsection{Final Focus R\&D studies}

The challenge of colliding nanometer-sized beams at the interaction point (IP) requires first creating very small emittance beams, then preserving this emittance during acceleration and transport, and finally focusing the beams to the nanometer scale vertically before colliding them.

The Accelerator Test Facility (ATF) at KEK is a prototype damping ring that has succeeded in achieving emittances that near the ILC requirements.  The ATF is being used as an injector for the ATF2, a special final-focus test beam line, which is serving as the test facility to follow-up the Final-Focus Test Beam (FFTB) studies at SLAC.  The ATF-2 has different beam-line optics, which are based on a scheme of local chromaticity correction that promise better performance and greater extendibility to higher energies.  

The primary goals of ATF2 are to achieve a 37 nm vertical beam size at the IP (equivalent to the ILC requirement) and to stabilize it at the nanometer level.  The optics of the ATF/ATF2  is a scaled-down version of the ILC design.  The beam line extends over about 90 meters from the beam extraction point from the ATF  damping ring to the final focus.

Anticipating gradual movements of supports and magnets due to thermal variations or slow ground motion, twenty quadrupole and five sextupole magnets are mounted on remote-controlled three-axis movers recycled from the FFTB experiment. The movers have a precision  1-2~$\mu$m for transverse motion (horizontal and vertical), and 3-5~$\mu$rad for rotations about the beam axis.

Overall alignment precisions of 0.1 mm (displacement) and 0.1 mrad (rotations) have been achieved using conventional alignment/metrology techniques. The final alignment of the magnets is achieved via beam-based alignment techniques.

\begin{figure}[h]
\centerline{\includegraphics[width=10.8cm]{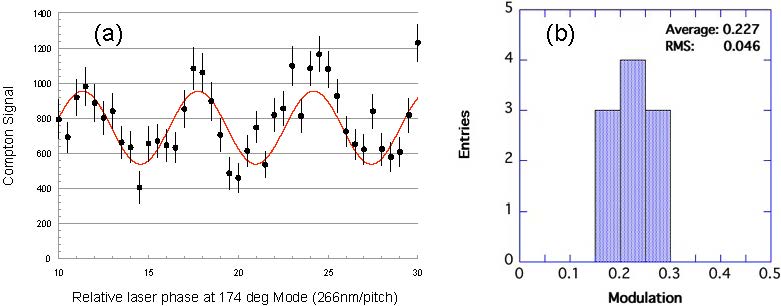}}
\caption{ATF2 measurements of the modulation corresponding to a beam size of $\sim70$ nm.
\label{f11}}
\end{figure}

The ATF/ATF2 program was set back about a year due to the Japanese earthquake in 2011, however the program is continuing toward achieving its goals.  In fact, recent results (summarized in Fig. \ref{f11}) are encouraging in that beam sizes within a factor of two of the goals has been achieved and no fundamental obstacles are anticipated in eventually achieving the ILC design requirements.

The ATF2 has achieved a $\sim 70$ nm beam size (a factor of $\times 2$ from the goal). This represents an important milestone towards verification of the local chromaticity correction scheme at the ILC final focus system. However, it should be pointed out that the results have only been obtained at very low beam intensity and further studies toward the 37 nm goal are proceeding.   Once the beam size goal is achieved, the R\&D program will focus on stability studies that are also crucial to obtaining design luminosity

%\end{document}

%% file: detectors.tex
The ILC physics program demands detectors with significant performance advances over past collider detectors.  With a machine environment that is benign by LHC standards, designs and technologies unthinkable at the LHC can be realized.  In spite of this advantage, there are special issues with the ILC environment that must be addressed.

High-resolution jet energy reconstruction and di-jet mass performance are among the critical requirements.  The particle flow algorithm (PFA) reconstruction technique has been developed to meet this challenge.  Highly granular electromagnetic and hadron calorimeters are deployed to achieve the needed jet energy resolution of 3 to 4 percent for 100 GeV jets, a goal set by the need to separate W and Z di-jet final states.  The charged track momentum resolution requirement is set by
the reconstruction of a recoiling Higgs boson associated with a Z boson decaying to a lepton pair.  The Higgs mass must be reconstructed from the lepton pair measurement, requiring a $\Delta p/p^2$ of better than $5 \times 10^{-5}  (GeV/c)^{-1}$.  Flavor and quark-charge tagging are
needed and provided by a new generation of vertex detectors. The highly granular calorimeters supply particle identification and muon detection is aided by instrumenting the iron return yoke.

These requirements far surpass the performance achieved in past and present collider detector experiments,
including the LHC. The specific conditions found at the ILC (particularly the beam time structure
and low radiation levels)
 make this advanced performance
possible, and the detector R\&D program undertaken in a globally coordinated effort have
developed the needed technologies.  As a result, the above performance parameters 
will provide the critical element to achieve the scientific goals of the ILC.
The two full detector design concepts of the ILC, SiD and ILD, build on these detector capabilities,
and are described below.

\subsection{Two detectors}

Concurrent operation of more than one experiment has been shown to be a critical element in 
scientific progress at recent and current colliders, such as the Tevatron, LEP, HERA, and the LHC. 
Each experiment brings complementary approaches, cross-checks for confirmation of results, 
reliability, insurance against mishaps, and competition. Additionally, this increases opportunities 
for personnel to participate in the science. Numerous historical examples demonstrate the importance of this, 
such as the inability of UA2 to confirm the mistaken claim for top quark discovery by UA1\cite{refUA2}.
Based on these considerations, the ILC has been designed to enable two 
experimental detectors with complementary features.

Through a validation process, two detailed detector designs have been developed to fulfill this important capability.
Both detector designs, SiD and ILD,
are conceived as multi-purpose detectors, optimized for
the broad range of physics opportunities at the ILC.
SiD is a compact, cost-constrained detector made possible with a 5 Tesla magnetic field
and silicon tracking.  Silicon enables time-stamping on single bunch crossings 
to provide robust performance, derived from immunity to spurious background bursts.
The highly granular calorimeter is optimised for particle flow analysis. 
The ILD group has designed a large detector with robust and stable performance over a wide range of energies. 
The concept uses a primary tracking system based on a continuous readout time projection chamber (TPC) 
to achieve excellent efficiency and robust pattern recognition performance. 
Some silicon tracking is employed to complement the TPC.
A granular calorimeter 
system contained inside a 3.5 T magnetic field provides very good particle flow reconstruction.  
The significantly different tracking technologies along with the compact (SiD) vs. large (ILD) 
designs achieves the needed, valuable detector complementarity.  
Both detectors provide flexibility for operation at energies up to the TeV range.

The two detectors share the one interaction region through a `push-pull' system.
This system allows
each detector be moved to the beamline to operate with collisions 
while the other detector is parked out of
the beam in a close-by maintenance position.  
The intervals of switching locations are
short enough to ensure both acquire data on any potential discovery. The time for transition must be
on the order of one day to maximize ILC integrated luminosity.

\begin{figure}[b]
\centerline{\includegraphics[width=8.8cm]{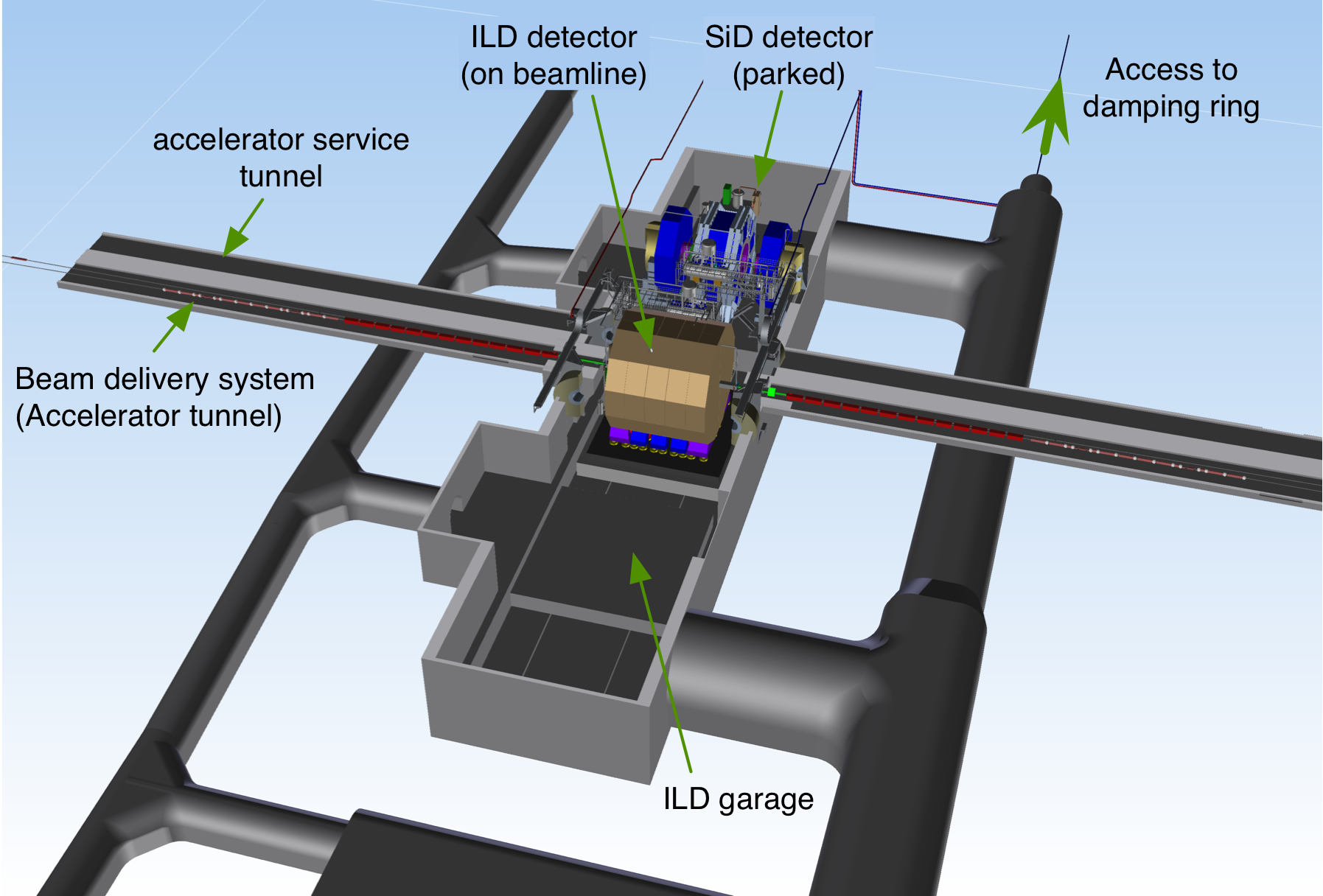}}
\caption{Layout of the detector hall for the Mountain
Topography site, showing the location of the two detectors,
when ILD in on the beamline and SiD is parked in the close-by maintenance position,
illustrating the `push-pull' arrangement. \label{two-fig}}
\end{figure}

The detectors and associated infrastructure are
arranged to enable quick movement of each detector into and out of the interaction region. 
This requires a time-efficient implementation of `push-pull', which sets 
requirements\cite{mdi} and challenges
for many detector and machine systems, in particular the IR magnets, the cryogenics, the
alignment system, the beamline shielding, the detector design, and the overall integration. 
Detailed designs of technical systems and the experimental area have been developed from 
the engineering specifications.
The layout of the detector hall is illustrated in Fig. \ref{two-fig}.
The detector motion and support systems are designed for roughly one hundred moves, with
preservation of detector internal components and detector positioning accuracy.
The motion system must also preserve
structural integrity of the collider hall floor and walls, be compatible with vibration stability of the
detector at the level of tens of nanometers and be compatible with earthquake-safety standards.
The alignment of detectors must be re-established after each movement.
The detectors will be placed on platforms that preserve the detector alignment and distribute
the load evenly onto the floor.  The ILC detectors are self-shielding with respect to
ionizing radiation from maximum credible beam-loss scenarios, although, while in the beam position,
added shielding must fill the gap between the detector and the wall.

The stray magnetic fields outside the iron return yokes of each detector must be less than 15 mT
at a lateral distance of 15 m from the beamline so as not to
disturb the other detector during operation or maintenance. 
Fringe fields from the detector return yokes
have been carefully simulated and designs for both SiD and ILD have been developed to meet these
requirements.

Since the European and American sample sites assume a flat surface area while the Asian
sample sites in Japan are located in mountains,
the requirements for conventional facilities and buildings are different, as are
the installation schemes for the detectors and the layout of the experimental areas.

\subsection{ILC physics environment and machine backgrounds}

At 250 GeV center-of-mass energy, the time structure of the ILC beams consists of 1312 particle bunches in a 724 $\mu$sec bunch train
delivered at 5 Hertz.  Within a bunch train the interbunch separation is 552 nsec. The detectors
must be designed with sensitivity to this time structure.

While benign by LHC standards,  ILC
machine backgrounds must be addressed for optimal performance of the detectors.
A variety of processes create beam induced backgrounds in the detectors. The main sources are beamstrahlung, synchrotron radiation,
muons, and neutrons.  
Significant numbers of low energy $e^+e^-$ pairs are produced at 
the interaction point from the beam collision induced beamstrahlung;
the interaction region layout has been designed to
guide these charged background particles out of the detector
by adding a dipole field to the conventional solenoidal field
(so-called {\it anti-DID}). These background pairs must be dealt with in the vertex detector design, but they
primarily affect the very forward detectors, which need to be able to withstand the significant radiation loads, and at the same time maintain sensitivity to single high energy particles. 
An optimized collimation system is used to control the upstream generated
synchrotron radiation.  Various schemes are employed to minimize the impact of 
muons and neutrons.

Another important background comes from photon-photon collisions. These events produce high-p$_T$ particles which overlap with the particles from the less common hard scattering events. Time stamping at the single bunch level can help in reducing the number of overlapping events. In addition sophisticated algorithms have been developed to identify and subtract these events based on topology and detailed properties.

\subsection{Detector R\&D}

The capabilities required for fully integrated experimental detectors
have been developed through a coordinated world-wide ILC detector R\&D
program.  The critical aspects of this program involve calorimetry, tracking,
and vertex sensors.  Since
many of the important physics channels for the ILC involve multi-jet final states,
calorimetry considerations begin by addressing this challenge.
Clean separation of jets, with reconstruction of the two jet invariant masses, is 
essential. The invariant mass resolution requirement is set by the gauge boson
widths of about 3\%. The PFA approach offers a promising method to
achieve this unprecedented mass resolution, and it has been adopted by the
ILC detectors. The four-vectors of all visible particles are reconstructed first,
followed by jet reconstruction. For charged tracks measured in the tracker the tracker
momentum measurement provides this four-vector.  This is superior to the measurement
of these charged tracks in the calorimeter.  Photons and neutral hadrons are measured
in the calorimeter, but this requires separation in the calorimeter of each of the
interacting particles.  In practice, perfect separation is not achievable and
``confusion" adds to the performance.  Optimizing this separation motivates highly
granular calorimeters, both longitudinally and transversely.

\begin{figure}[b]
\centerline{\includegraphics[width=8.8cm]{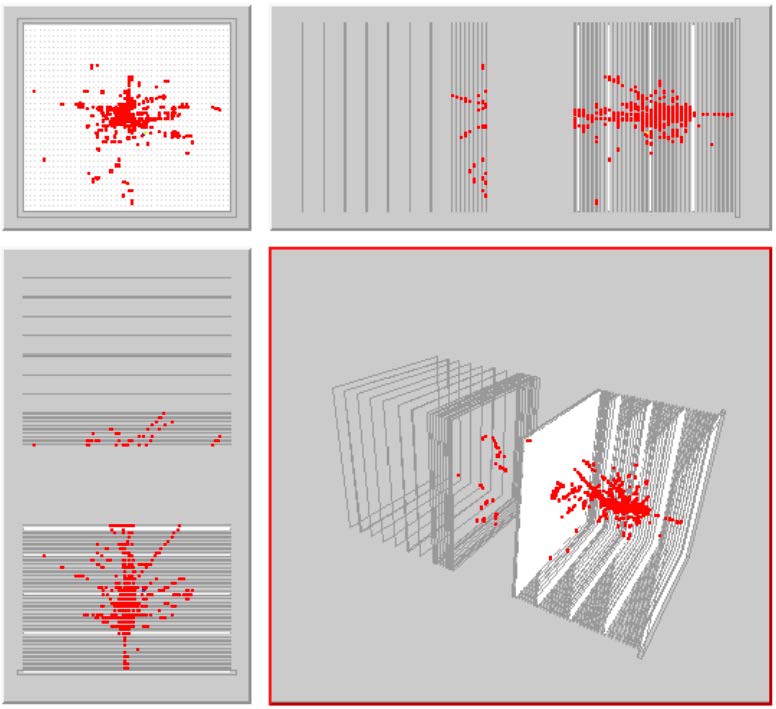}}
\caption{Event display showing the interaction of a 10 GeV pion in the
CALICE DHCAL with RPC read-out. \label{dhcal-rpc}}
\end{figure}

New technology development has been a focus of detector R\&D for calorimetry. Both tungsten 
and iron have been studied for hadron calorimeter (HCAL) radiators, and many
active technologies are being pursued, including 
scintillators with novel SiPM readout, silicon based monolithic
active pixel sensors (MAPS) and gaseous detectors such as Resistive Plate Chambers (RPCs) and
Micromegas. 
A broad R\&D effort has been carried out to test these technologies and validate the simulations
and the PFA performance predictions, largely conducted by the CALICE Collaboration.
For example, a digital HCAL prototype instrumented with glass RPCs containing nearly
500,000 channels was tested yielding highly detailed event images such as that 
illustrated in  Fig. \ref{dhcal-rpc}.

The electromagnetic calorimeter must
identify photons, measure their energy,
and separate the photons from each other and from near-by hadrons
for the PFA jet reconstruction.
Tungsten is a good choice for the radiator due to the
large difference between electromagnetic
radiation length and nuclear interaction length; it also has a small Moliere radius,
which minimizes the transverse spread of electromagnetic showers.
Silicon pad diodes (or monolithic active pixels) offer the highest
possible compactness (and effective Moliere radius) with excellent stability of calibration.
Scintillating strips with silicon photo-detector readout have a similar effective segmentation at a lower cost, but are less compact.
Each of these options are possible to be combined in a hybrid system.

Tracking technology developments are both silicon- and gaseous-based. For the former, silicon
strip technology is being pursued as well as the option of  highly pixelated silicon sensors.
The gaseous tracking approach is based on a Time Projection Chamber (TPC), including studies
of Micromegas, GEM and CMOS devices to collect drifted charge.

Heavy fermion detection relies on topological reconstruction of displaced vertices.
Detection of such displaced vertices relies an single point resolution of vertex sensors 
near the interaction point (IP). Fine pitch, low-mass pixel detectors located as close
to the IP as possible are needed. 
The first sensor layers of the ILC detectors are located
14-16 mm from the IP.
The allowed
material budget  is about 0.1\% X$_0$ per layer for the vertex detector and less than 1\% X$_0$ per
layer for a silicon tracker. For a TPC the material budget is accumulated in the end plates and a
material budget of 30\% X$_0$ per endplate is the goal. 

To achieve an ultra low-mass detector configuration, the low duty cycle of the ILC machine is
exploited. The detector can thus be put in a quiescent state, the ``power-pulsing" mode, 
for about 199 out of 200 milliseconds at the
machine repetition rate of 5 Hz, leading to an overall reduction in power consumption of nearly a
factor of hundred. One of the most significant
benefits of ``power-pulsing" is that the vertex and tracking detectors do not need active cooling,
lowering the overall mass budget for these detectors, which is crucial for obtaining the
required resolutions. A 20 W heat load for the barrel vertex detector is thought to be removable
using forced convection with dry air.

\subsection{Detectors: SiD}

\begin{figure}[b]
\centerline{\includegraphics[width=8.8cm]{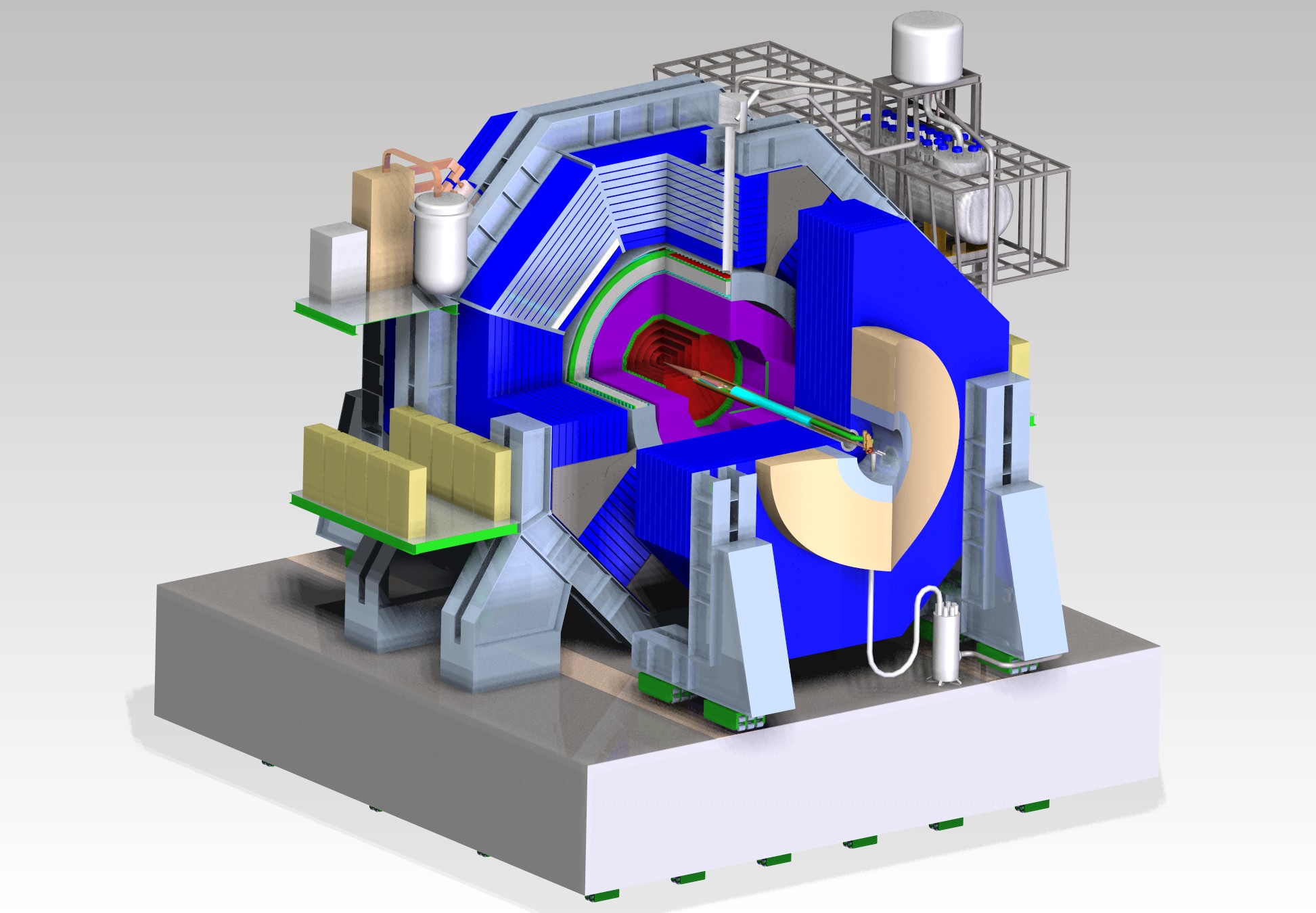}}
\caption{An isometric view of SiD on its platform, showing tracking (red), ECAL (green), HCAL (violet)
and the flux return (blue). \label{sid-fig}}
\end{figure}

SiD is a general-purpose detector designed to satisfies the challenging detector requirements of the ILC
with the capability to perform precision measurements.
It was conceived from its inception as a fully integrated, unified design, based on a compact silicon tracking based
detector, with a finely grained calorimeter inside a high central solenoidal magnetic field.
It results from many years of creative design by physicists and engineers, backed up by a substantial body of past and ongoing detector R\&D. While each component has benefitted from continual development, the SiD design integrates these components into a complete system for excellent measurements of jet energies, based on the PFA approach, as well as of charged leptons, photons and missing energy. The use of robust silicon vertexing and tracking makes SiD applicable to a wide range of energies from a Higgs factory to beyond 1 TeV. SiD has been designed in a cost-conscious manner, with the compact design that minimizes the volumes of high-performing, high-value, components, while maintaining critical levels of performance. The restriction on dimensions is offset by the relatively high central magnetic field from a superconducting solenoid.
An image of SiD is shown in Fig. \ref{sid-fig}. The radius of SiD is 604 cm
and the length is 1135 cm.

The use of silicon detectors for tracking and vertexing ensures SiD robustness with respect to beam backgrounds or beam loss, provides superior charged-particle momentum resolution, and eliminates out-of-time tracks and backgrounds. The main tracking detector and calorimeters are only {\it live}  during each single bunch crossing, so beam-related backgrounds and low-p$_T$  backgrounds from $\gamma\gamma\rightarrow$ hadrons  processes will be reduced to the minimum possible levels. The SiD calorimetry is optimized for excellent jet-energy measurement using the PFA technique. The complete tracking and calorimeter systems are contained within a superconducting solenoid, which has a 5 T field strength, enabling the overall compact design. The coil is located within a layered iron structure that returns the magnetic flux and is instrumented to allow the identification of muons.

The PFA 
relies on a high resolution tracking system with
superb efficiency and good two-particle separation. The physics requirements, in particular in the Higgs sector, place high demands on the momentum resolution at the level of $\delta(1/p_T) \sim 2-5 \times 10^{-5} (GeV/c)^{-1}$  and on the material budget of the tracking system. Highly efficient tracking is achieved using the pixel detector and main tracker to recognize and measure prompt tracks.

The SiD vertex detector layout includes a barrel and a forward disk system. The barrel section consists of five silicon pixel layers with a pixel size of $20 \times 20 ~ \mu$m$^2$ with radii ranging from 14 to 60 mm.
 The forward and backward regions each have four silicon pixel disks. In addition, there are three silicon pixel disks at a larger distance from the IP to provide uniform coverage for the transition region between the vertex detector and the outer tracker. This configuration provides for very good hermeticity with uniform coverage and guarantees excellent charged-track pattern-recognition capability and impact-parameter resolution over the full solid angle. The vertex detector design relies on ``power pulsing" during bunch trains to minimize heating and uses forced air for its cooling. The main tracker technology of choice is silicon-strip sensors arrayed in five nested cylinders in the central region with an outer cylinder radius of 1.25 m and four disks in each of the endcap regions. The geometry of the endcaps minimizes the material budget to enhance forward tracking. The detectors are single-sided silicon sensors with a readout pitch of 50 $\mu$m.

PFA operation constrains the calorimeter design.  The central calorimeter system must be contained within the solenoid in order to reliably associate tracks to energy deposits. The electromagnetic (ECAL) and hadronic (HCAL) sections must have imaging capabilities that allow both efficient track-following and correct assignment of energy clusters to tracks. These requirements imply that the calorimeters must be finely segmented both longitudinally and transversely.

A central barrel with two endcaps, nested inside the barrel, comprise the calorimeter (ECAL and HCAL) system.
The entire barrel system is contained within the volume of the cylindrical superconducting solenoid. The electromagnetic calorimeter has silicon active layers between tungsten absorber layers. The active layers use $3.5 \times 3.5$ mm$^2$  hexagonal silicon pixels, which provide excellent spatial resolution. The structure has 30 layers in total, the first 20 layers having a thinner absorber than the last 10 layers. This configuration is a compromise between cost, electromagnetic shower radius, sampling frequency, and shower containment. The total depth of the electromagnetic calorimeter is
26 radiation lengths ($X_0$) and one nuclear interaction length. The hadron calorimeter has a depth of 4.5 nuclear interaction lengths, consisting of alternating steel plates and active layers. The baseline choice for the active layers is the glass resistive-plate chamber (RPC) with an individual readout segmentation of $10 \times 10$ mm$^2$. Two special 
luminosity calorimeters
are planned for the very forward region: LumiCal for precise measurement, and BeamCal for fast estimation.

CMS solenoid design philosophy and construction techniques
have been adopted for the SiD superconducting solenoid.
A slightly modified CMS conductor provides the SiD  baseline design. Superconducting strand count in the coextruded Rutherford cable was increased from 32 to 40 to accommodate the higher 5 T central field. The flux-return yoke is instrumented with position sensitive detectors to serve as both a muon filter and a tail catcher. The SiD Muon System baseline design is based on scintillator technology, using extruded scintillator readout with wavelength-shifting fibre and 
silicon photomultipliers (SiPMs). 
Simulation studies have shown that nine or more layers of sensitive detectors yield adequate energy measurements and good muon-detection efficiency and purity.

\subsection{Detectors: ILD}
 
\begin{figure}[b]
\centerline{\includegraphics[width=8.8cm]{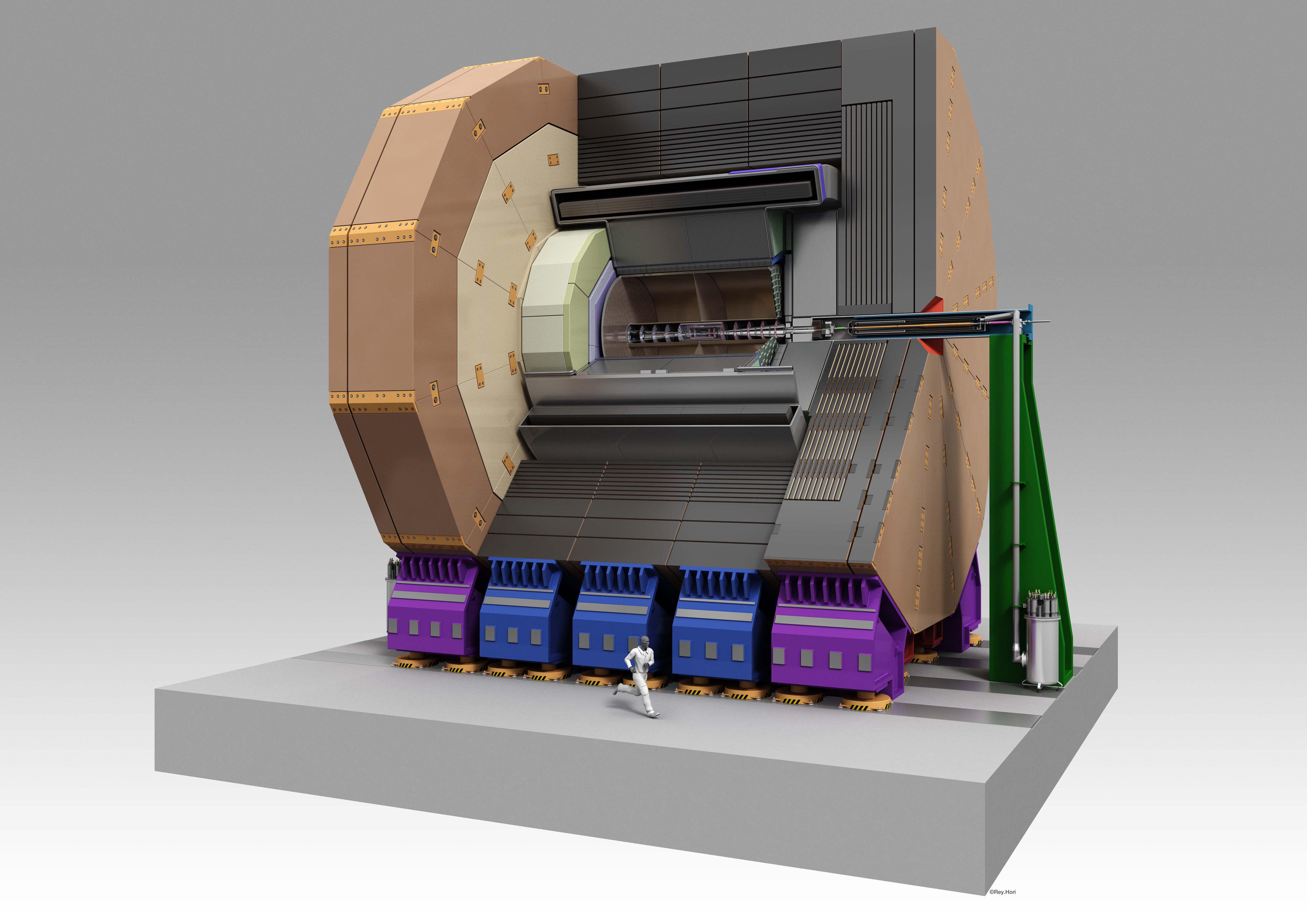}}
\caption{The ILD detector concept. \label{ild-fig}}
\end{figure}

ILD is a multi-purpose detector, with optimal PFA performance,  employing a high-precision vertex detector followed by a hybrid tracking system, realized as a combination of silicon tracking with a time-projection chamber (TPC), and a calorimeter system. The complete system is located inside a 3.5T solenoid. The high level of granularity of the inner-detector system provides a robust and detailed three-dimensional imaging capability. Outside of the coil, the iron return yoke is instrumented as a muon system and as a tail-catcher calorimeter. A view of the detector is shown in Fig. \ref{ild-fig}. The radius of ILD is 783 cm
and the length is 1324 cm.

The multi-layer ILD pixel vertex detector (VTX) has three superlayers each comprising two layers, or,
optionally,  five single layers. In either case the detector has a pure barrel geometry with radii ranging from 16 to 60 mm.
The first superlayer is only half as long as the outer two in order to
minimize occupancy from background hits. 
The underlying detector technology has not yet been decided, but the VTX is
optimized for point resolution and minimum material thickness.
A silicon strip and pixel detector system surrounds the VTX. In the barrel, two layers of silicon strip detectors (SIT) are arranged to bridge the gap between the VTX and the TPC. In the forward region, a system of two silicon-pixel disks and five silicon-strip disks (FTD) provides low angle tracking coverage.
A distinct feature of ILD is a large-volume TPC with up to 224 points per track. The TPC is optimized for 3-dimensional point resolution and minimum material in the field cage and the end-plate. 
It provides dE/dx -based particle identification.
Outside the TPC a system of silicon strip detectors provide additional high-precision space points, improving the tracking performance and providing additional redundancy in the regions between the main tracking volume and the calorimeters.
One  layer of detectors sits behind the end-plate of the TPC (ETD) and one between the TPC and the ECAL (SET), 

A highly segmented electromagnetic calorimeter (ECAL) provides up to 30 samples in depth with a small transverse cell size.
The ECAL is split into a barrel and an endcap system. The absorber is tungsten; silicon diodes, scintillator strips or a combination are under consideration for the sensitive layers.
The ECAL is followed by a highly segmented hadron calorimeter (HCAL) with up to 48 longitudinal samples and small transverse cell sizes. Two options are under consideration, both based on a steel absorber structure. One option uses scintillator tiles of $3 \times 3$  cm$^2$, which are read out with an analogue system. The second uses a gas-based readout which allows a $1 \times 1$ cm$^2$  cell geometry with a binary or semi-digital readout of each cell.
At very forward angles, below the coverage provided by the ECAL and the HCAL, a system of high-precision and radiation-hard calorimetric detectors (LumiCAL, BeamCAL, LHCAL) is foreseen. These extend the calorimetric solid-angle coverage to almost 4$\pi$, measure the luminosity, and monitor
the quality of the colliding beams.

The ILD calorimeters are surrounded by a
large volume superconducting coil, creating an axial B-field of nominally 3.5 Tesla. An iron yoke, instrumented with scintillator strips or resistive plate chambers (RPCs), returns the magnetic flux of the solenoid, and also serves as a muon filter, muon detector and tail-catcher calorimeter. 

To maximize the sensitivity of ILD to ILC physics, it will be operated in a continuous readout mode, without a traditional hardware based trigger. 

In close collaboration with the detector R\&D groups, all key components of the ILD proposal have been evaluated and the key performance criteria have been demonstrated in test-beam experiments.
This is of particular importance for the very ambitious calorimeter proposal.

A first engineering study of the integration of the ILD detector has been performed. This study has included a detailed review of the detector components, their sizes, and in particular the support structures needed to mount the detector. Assembly and maintenance procedures have been simulated to validate the integration concept. Estimates of services needed have been included as far as possible, and realistic tolerances have been inserted into the designs. A detailed model of the complete detector has been built using modern CAD tools, and compared to the simulation model
used for the performance evaluation.

\subsection{Detector performance}

\begin{figure}[htbp]
\parbox{2.4in}{
\begin{center}
\includegraphics[width=2.4in]{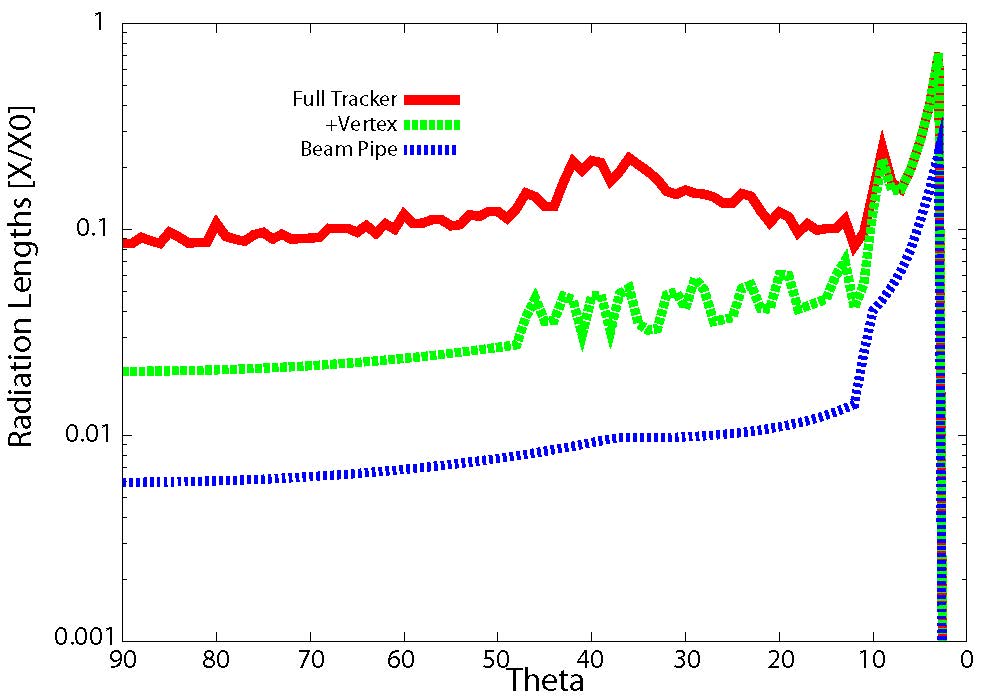}
\end{center}
%\caption{ 
%}
}~~~~~\parbox{2.8in}{
\begin{center}
\includegraphics[width=2.4in]{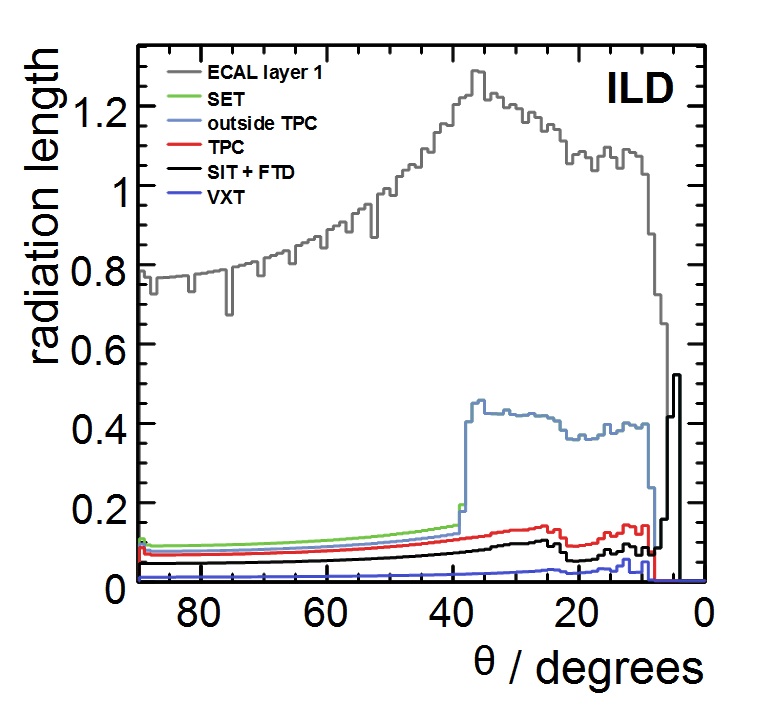}
\end{center}
%\caption{ }
%}
}
\caption{Average total radiation length of the material
  in the tracking detectors as a function of polar angle for SiD (left) and ILD(right).
  The red curve on the ILD figure shows the material in the tracking system without the outer material in the field cage or endplate; the grey line shows the total material up to the first active layer inside the ECAL is reached.}
\label{trk-material}
\end{figure}

The two detectors share
a large fraction of the software for the generation, simulation and reconstruction. 
The SiD detector concept is fully implemented and simulated in the GEANT4-based SLIC software package. 
Likewise, the ILD concept is modeled in detail with GEANT4.  
The performance of the both detectors have been extensively studied with these
detailed simulation models and sophisticated reconstruction tools. 
Backgrounds, such as
incoherent pair interactions and $\gamma \gamma \rightarrow$ hadrons for one bunch crossing,
have been carefully taken into account.  
Events are then passed through the reconstruction software suite, encompassing digitization, tracking, vertexing and the PandoraPFA algorithm.\cite{pfa} 
The material with each detector is a key characteristic. 
PFA calorimetry requires a thin tracker, to minimize interactions before the calorimeters, and thick calorimeters, 
to fully absorb the showers. 
The tracker material budget for SiD is less than 20\% $X_0$ down to very low angles and the amount of tracker material in ILD up to the end of the tracking volume is mostly below 10\% $X_0$. 
Fig. \ref{trk-material} shows the material in the SiD (left) and ILD (right) tracking detectors in radiation lengths ($X_0$). 

%The right-hand plot shows the total interaction length including the calorimeter system. 

%\begin{figure}[b]
%\centerline{\includegraphics[width=8.8cm]{images/detectors/ES/sidloi3TrackerMaterialScan.pdf}}
%\caption{The ILD detector concept. \label{trk-material}}
%\end{figure}

The increase of TPC material in the ILD endcap due to the support structure for the readout in the TPC
is clearly visible in right side of Fig. \ref{trk-material}.
Since this material is located very close to the endcap calorimeter it has 
a very small negative impact on the performance, as can be seen from the top curve in the plot, 
the total number of radiation length until the first active layer inside the ECAL.

\begin{figure}[htbp]
\parbox{2.5in}{
\begin{center}
\includegraphics[width=2.5in]{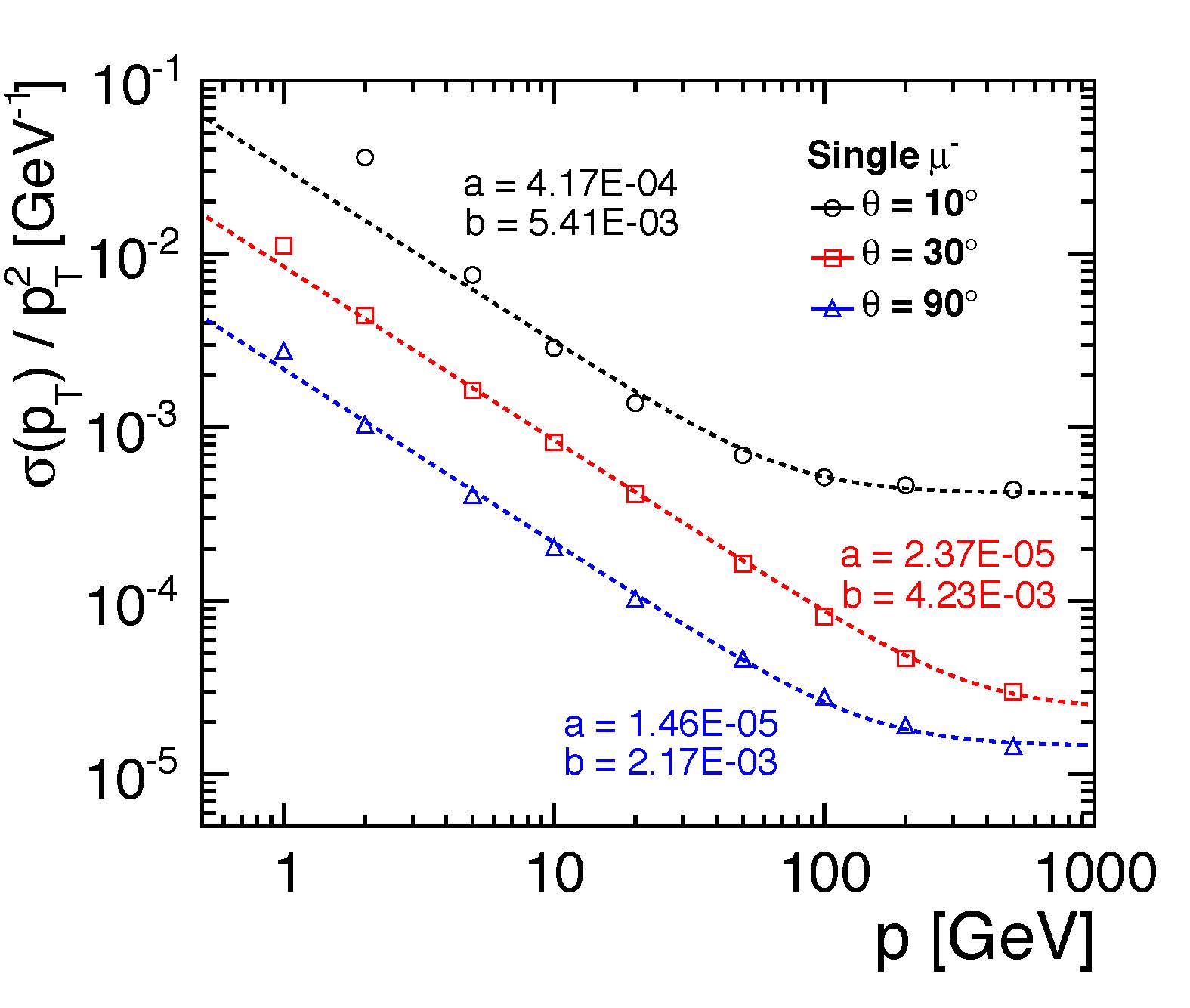}
\end{center}
%\caption{ 
%}
}~~~~~\parbox{2.5in}{
\begin{center}
\includegraphics[width=2.5in]{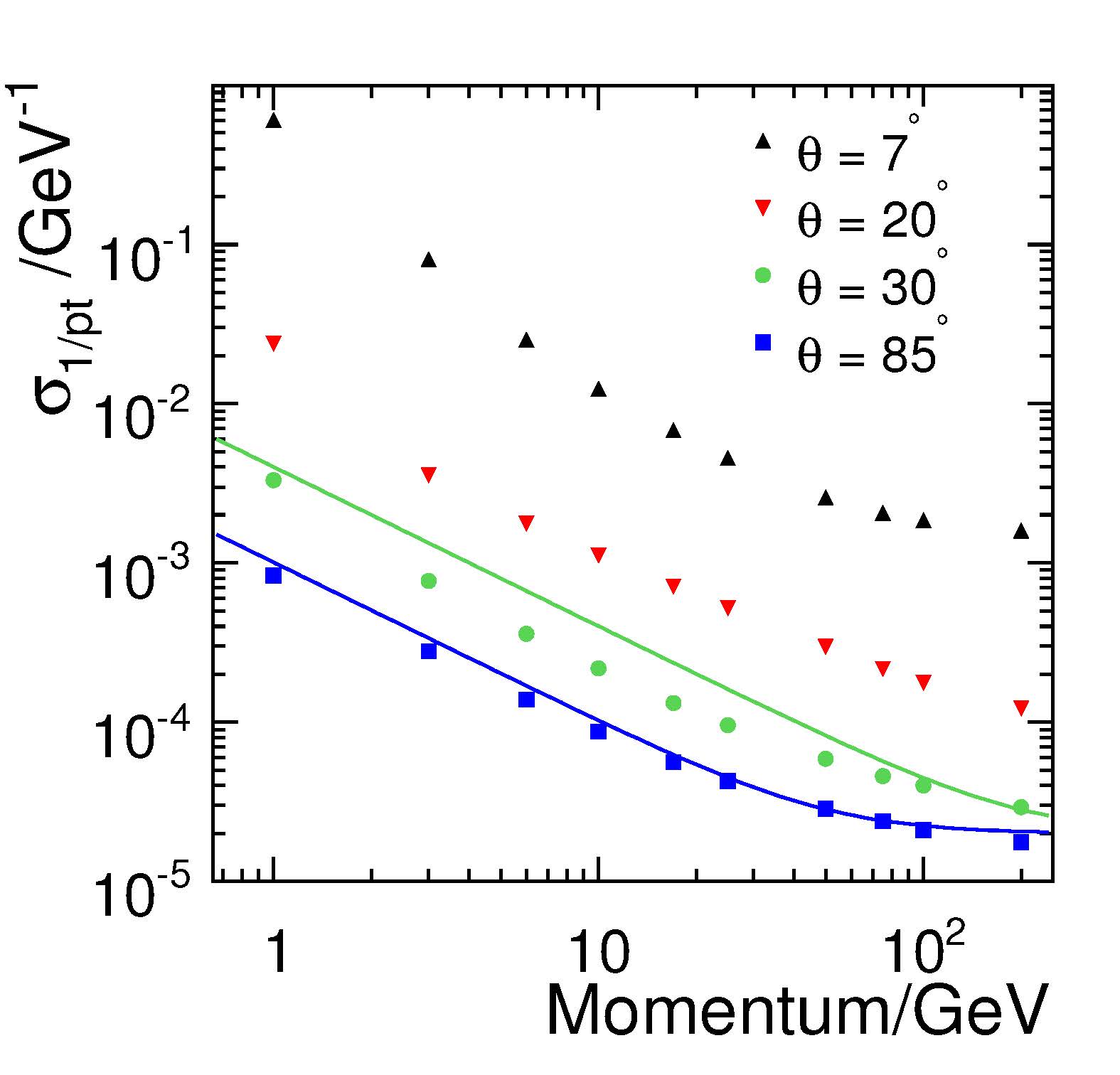}
\end{center}
%\caption{ }
%}
}
\caption{Momentum resolution as a function of the transverse momentum of particles, for tracks with different polar angles
for SiD(left) and ILD (right). Also shown for SiD is the parametrization for the constants shown, and for ILD the theoretical expectation.}
\label{trk-perf}
\end{figure}

The simulated tracking performances for single particles are shown in Fig. \ref{trk-perf}.
The SiD design achieves an asymptotic momentum resolution of $\delta(1/p_T) = 1.46 \times 10^{-5}$ (GeV/c)$^{-1}$, while the ILD design high momentum resolution is $\sigma_{1/p_T} = 2 \times 10^{-5}$ (GeV/c)$^{-1}$.

\begin{figure}[htbp]
\parbox{2.5in}{
\begin{center}
\includegraphics[width=2.5in]{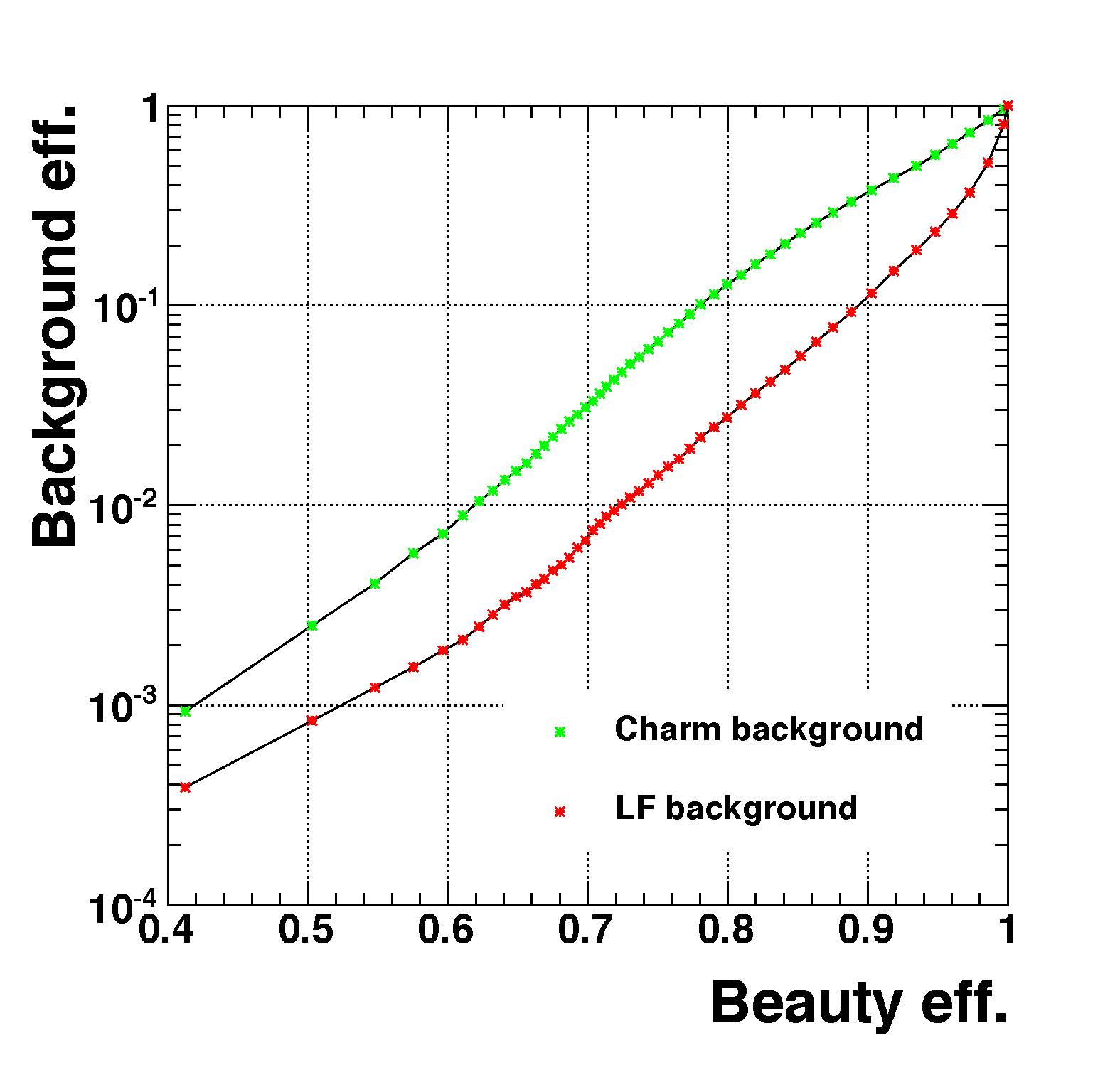}
\end{center}
\caption{ SiD mis-identification efficiency of light quark (red points) and
c quark events (green points) as b quark jets versus the b
identification efficiency in di-jet events at \roots = 91~GeV including background 
from \gghadrons and incoherent pairs.
}
\label{sid-b}
}~~~~~\parbox{2.5in}{
\begin{center}
\includegraphics[width=2.5in]{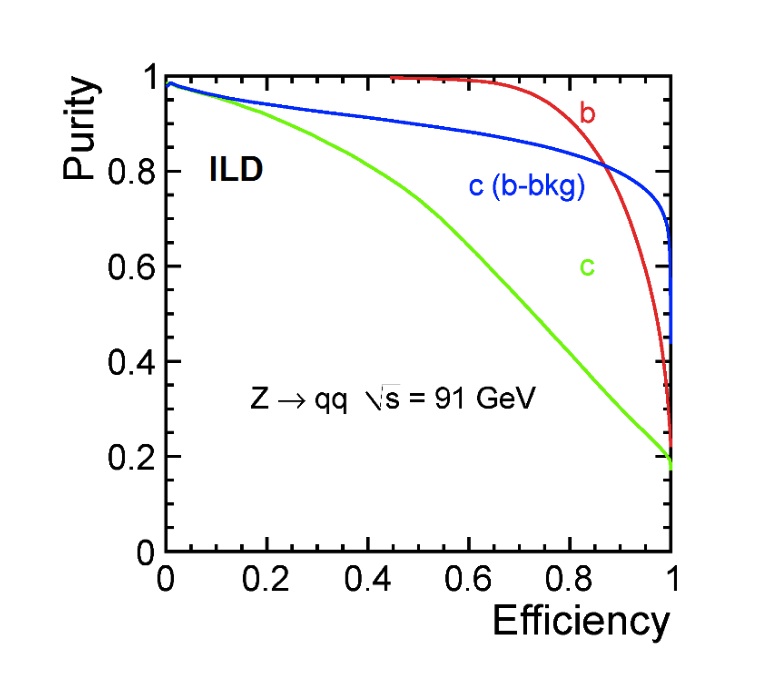}
\end{center}
%\caption{ }
%}
\caption{Flavour tagging performance of ILD for
    $Z\rightarrow q\overline{q}$ samples
    at 91~GeV.}
\label{ild-vert}
}
\end{figure}

For many physics studies the tagging of long-lived particles is of key importance. The ability to tag bottom and charm decays with high purity has been a driving force in the design of the vertex detectors.
With a transverse impact parameter resolution of better than 2 $\mu$m, Fig. \ref{sid-b} shows SiD's ability to separate b-quarks also in the presence of the full beam background. The ability of ILD to reconstruct displaced vertices is shown in Fig. \ref{ild-vert}.

\begin{figure}[htbp]
\parbox{2.5in}{
\begin{center}
\includegraphics[width=2.5in]{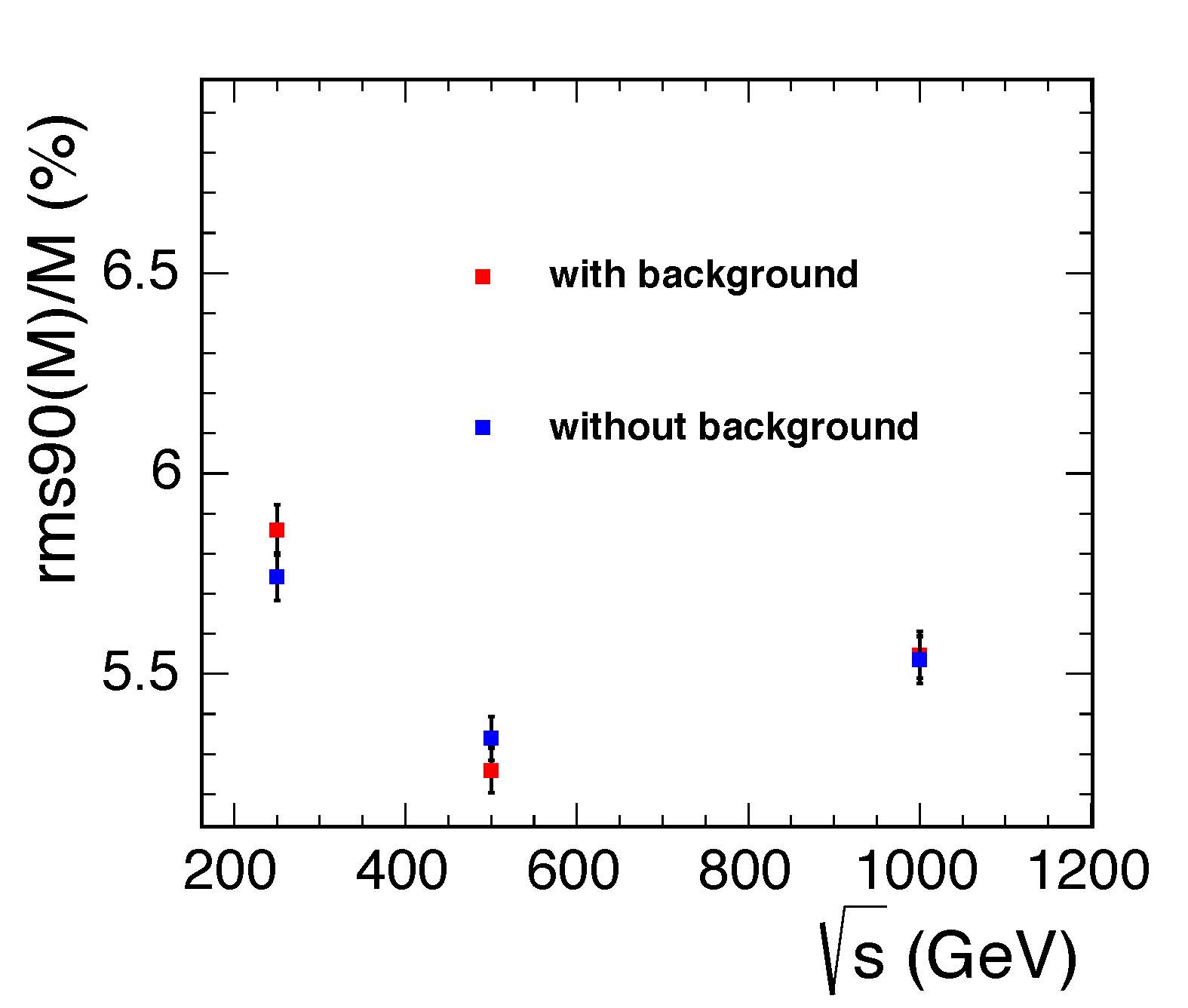}
\end{center}
%\caption{ 
%}
}~~~~~\parbox{2.5in}{
%\begin{center}
\includegraphics[width=2.5in]{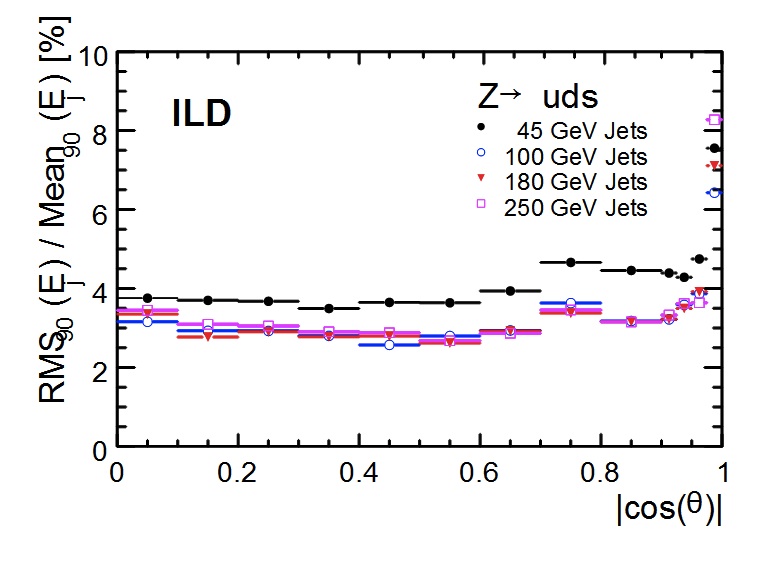}
%\end{center}
%\caption{ }
%}
}
\caption{Left: SiD mass resolution of reconstructed $\PZ\PZ$ events with and without the backgrounds from
$\gamma\gamma\rightarrow$hadrons and incoherent pairs at different values of $\sqrt{s}$.
Right: Fractional jet energy resolution in ILD
    plotted against $|\cos\theta|$ where $\theta$ is the polar angle of the thrust axis of the event.
}
\label{cal-res}
\end{figure}

Jet reconstruction relies on excellent tracking and calorimetry, combined in sophisticated reconstruction algorithms based on the PFA technique. The excellent resolution achieved this way allows the separation of W and Z decays.  The simulated jet-energy resolution performance of the SiD and ILD detectors are shown in Fig. \ref{cal-res}.
The SiD performance (left) is shown as the mass resolution of a Z boson decaying to two jets
in $e^+e^- \rightarrow ZZ$ where the other Z decays to two neutrinos. The right side of Fig. \ref{cal-res} shows
the ILD fractional jet energy resolution as a function of polar angle for different jet energies.
 (Resolutions are presented as the standard deviation in the smallest range that contains 90\% of the events, so-called rms90. This measure avoids the impact of tails in the energy distribution.)

\subsection{Beam instrumentation}

The ILC physics program relies on
precise knowledge of beam parameters.
Instrumentation close to the main detectors measures
instantaneous luminosity, beam energy, and polarization.
The luminosity determination comes from
low-angle Bhabha scattering detected by dedicated calorimeters. Acollinearity
and energy measurements of Bhabha events in the polar angle region from
120-400 mrad reveals the luminosity as a function of energy, dL/dE, with the necessary precision.
Both upstream and downstream 
beam energy measurements with an accuracy of (100-200) parts per million 
provide redundancy and reliability of the results. 
Dedicated Compton polarimeters detect backscattered electrons and
positrons to provide the required precision,
0.25\% or better for
precise measurements of parity-violating asymmetries 
and 0.1\% precision for high statistics Giga-Z running.
Polarimeters implemented both upstream and downstream of the 
interaction region achieves the
best accuracy.

%% file: conclusion.tex
The ILC and the detector designs described here have the capability for an unprecedented
physics program in electron-positron collisions from a center-of-mass energy of 250 GeV
up to 1 TeV.  Beginning with a detailed study of the 125 GeV Higgs boson, the ILC should
significantly advance elementary particle physics.

%% file: acknow.tex
The authors thank the International Committee on Future Accelerators (ICFA)
and the International Linear Collider Steering Committee (ILCSC) for 
scientific and technical guidance, the Funding Agencies for the Linear Collider (FALC)
for international funding agency guidance, and the US Department of Energy and
the US National Science Foundation for their support.
We thank our many colleagues whose work is summarized here,
particularly the members of the Global Design Effort (GDE) for
carrying out the accelerator R\&D and design and the GDE Executive Committee
for providing overall technical leadership. We also acknowledge our
many colleagues of the ILC Research Directorate,
the SiD and ILD detector collaborations, the ILC detector R\&D collaborations
 and the World Wide Study for Future
electron-positron Linear Colliders, for developing the scientific goals,
studying the reach for new physics, designing the detectors and carrying
out the detector R\&D that proved the detector technologies.